\documentclass[%
reprint,
 groupedaddress,
 amsmath,amssymb,
 aps,
 prd,
]{revtex4-1}

\usepackage{graphicx}% Include figure files
\usepackage[utf8]{inputenc}
\usepackage{hyperref}% add hypertext capabilities
\usepackage{color}

\newcommand{\Mc}{M_{\rm c}}
\newcommand{\Mh}{M_{\rm h}}

\begin{document}

\title{Simulations of solitonic core mergers in ultra-light axion dark matter cosmologies}

\author{Bodo Schwabe}
\author{Jens C. Niemeyer}
\author{Jan F. Engels}
%\email{}
\affiliation{%
 Institut f\"ur Astrophysik\\
 Universit\"at G\"ottingen
}%

\date{\today}

\begin{abstract}
 Using three-dimensional simulations, we study the dynamics and final structure of merging solitonic cores predicted to form in ultra-light axion dark matter halos. The classical, Newtonian equations of motion of a self-gravitating scalar field are described by the Schrödinger-Poisson equations. We investigate mergers of ground state (boson star) configurations with varying mass ratios, relative phases, orbital angular momenta and initial separation with the primary goal to understand the mass loss of the emerging core by gravitational cooling. Previous results showing that the final density profiles have solitonic cores and NFW-like tails are confirmed. In binary mergers, the final core mass does not depend on initial phase difference or angular momentum and only depends on mass ratio, total initial mass, and total energy of the system. For non-zero angular momenta, the otherwise spherical cores become rotating ellipsoids. The results for mergers of multiple cores are qualitatively identical.
\end{abstract}

\pacs{Valid PACS appear here}
\maketitle

\section{\label{sec:intro} Introduction}

If dark matter consists of a cold, ultra-light (pseudo)scalar field, the statistical distribution and structural properties of collapsed halos may be modified with respect to the predictions of standard cold dark matter (CDM) \cite{Marsh2010,Marsh2013,Bozek2015,Marsh2015b}. String theory compactifications provide a class of well-motivated candidate particles with axion-like properties which are naturally ultra-light, so-called ultra-light axions (ULAs) \cite{Arvanitaki2010}. Observations that are sensitive to the small-scale structure of dark matter halos thus open a unique window onto fundamental physics. 

Roughly speaking, Heisenberg's uncertainty relation suppresses gravitational collapse on scales below the de Broglie wavelength of particles with virial velocities \cite{Hu2000,Woo2009}. For halo masses of $10^8$ $M_\odot$ at $z \sim 5$, this is of the order of kpc if the particle mass is $m \sim 10^{-22}$ eV \cite{Chavanis2011}. The strongest constraint in this mass range to date follows from the predicted suppression of early galaxy formation and the measured optical depth to reionization, yielding  $m > 10^{-22}$ eV \cite{Bozek2015}. While ULAs provide a well-motivated class of candidate particles, the same phenomenology applies more generally to any massive scalar field with negligible self-interactions in a coherent non-thermal state, e.g. as a consequence of being produced by vacuum realignment, which include subclasses of scalar field dark matter (SFDM) \cite{Matos2011,Matos2014,Matos2015} and Bose-Einstein condensate (BEC) dark matter \cite{Frieman1995,Matos2000,Suarez2014,Lee2016}. The class of scalars with negligible self-interactions in the mass range relevant for constraints from structure formation is often referred to as fuzzy dark matter (FDM) \cite{Hu2000}. We will follow this convention here.

To clearly distinguish the effects of FDM from other modifications of CDM with suppressed small-scale growth such as warm dark matter, it will be necessary to search for generic signatures of FDM on halo density profiles and substructure. One such signature may be the existence of compact solitonic cores embedded in a halo with NFW-like density profile, found in cosmological simulations that used the comoving Schrödinger-Poisson (SP) equations to model FDM \cite{Schive2014a} (who used the notation $\Psi$DM to emphasize the wavelike nature of dark matter). Their density profiles are governed by well-known equilibrium solutions for self-gravitating scalar fields in the nonrelativistic regime whose properties have been studied intensively in the context of Newtonian boson stars \cite{Ruffini,Guzman2004,Liebling2012} and BEC dark matter \cite{Baldeschi1983,Sin1994,RindlerDaller2012,Paredes2016}. The core mass obeys a scaling relation with the mass of the host halo, $\Mc \sim \Mh^{1/3}$, which can be motivated by identifying the characteristic scale height of the solitonic core with the virial velocity de Broglie wavelength \cite{Schive2014b}. The presence of solitonic cores was used to fit the profiles of dwarf galaxies and suggested as a possible solution of the cusp-core problem in CDM cosmologies \cite{Marsh2015b,Calabrese2016}. In collisions of solitonic cores with exact phase opposition, destructive interference gives rise to a short-range repulsive force between the cores \cite{Paredes2016}. As the authors of \cite{Paredes2016} pointed out, in the context of galaxy cluster observations with indications of an offset between dark and stellar matter \cite{Carrasco2010,Massey2015}, this effect can provide an alternative explanation to self-interacting dark matter. 

These results suggest a rich phenomenology of hierarchical structure formation in the presence of solitonic halo cores. They motivate an investigation of their potential impact on halo substructure, baryonic physics, and the properties of the earliest generation of galaxies. Ideally, this would be achieved by direct simulations of the SP equations in large cosmological boxes. Requirements on the spatial resolution of the SP equations, however, limit the box size of currently affordable simulations to  $\sim 1$ Mpc$^3$ in the interesting range of $m \sim 10^{-22}$ eV \cite{Schive2014a}, making a detailed study of halo and subhalo core mergers infeasible.

On the other hand, since the time scales for major mergers and subhalo evolution (determined by dynamical friction) are large compared to the gravitational time scales of the cores, much can be learned from the simplified problem of isolated mergers of two cores with different characteristic properties. One of the key questions is the efficiency of gravitational cooling of the newly formed core to shed mass and angular momentum as a function of the binary parameters \cite{Seidel1994,Guzman2006}. The results can be used, for instance, in semi-analytic models for galaxy formation in FDM cosmologies. 

In this work, we therefore address the simplified setup of merging solitonic cores in three-dimensional SP simulations. This allows us to perform parameter studies of total energy, mass ratio, angular momenta, and relative phase in order to map out their impact on the merging time, final core mass, and final angular momentum. Collisions and mergers of Newtonian boson stars have been studied extensively in 2D \cite{Bernal2006,Gonzalez2011} but we are unaware of fully three-dimensional simulations with no imposed symmetries. Furthermore, a systematic investigation of final core masses after relaxation by gravitational cooling has been lacking so far. In addition to studying binary mergers, we also relax the assumption of isolated events by considering mergers of multiple cores in fast succession in order to compare our results with those presented in \cite{Schive2014b}.

The remainder of this paper is structured as follows. In \autoref{sec:theory} we briefly outline the underlying theory. Our numerical methods are described in \autoref{sec:numerics}. In \autoref{sec:binary} we summarize results from an in depth analysis of binary mergers of two solitonic cores. In \autoref{sec:multimerge} we extend this investigation to mergers of multiple cores. We conclude in \autoref{sec:conclusion}.

\section{Solitonic halo cores}
\label{sec:theory} 

The nonrelativistic dynamics of a coherent massive scalar field can be described by a function $\psi$ which is governed by the Schr\"odinger-Poisson (SP) equations
\begin{align}
\label{eq:1}
  i\hbar\frac{\partial\psi}{\partial t} &= -\frac{\hbar^{2}}{2m}\nabla^{2}\psi+mU\psi\\
  \nabla^{2}U &= 4\pi G\rho\,\,,
\end{align}
where the density is defined as $\rho=|\psi|^{2}$. Simulations of the (comoving) SP equations with cosmological initial conditions show the formation of halo cores with solitonic profiles \cite{Schive2014a} that coincide with  spherically symmetric, stationary solutions of the SP equations otherwise known as (nonrelativistic) boson stars\footnote{To emphasize that this work is focused on galactic rather than stellar scales, we will refer to these solutions as (solitonic) cores instead of boson stars. All of our results, however, are independent of this interpretation.} \cite{Guzman2004, Liebling2012}. Their radial density profile is well approximated by \cite{Schive2014b}
\begin{align}
    \label{eq:2}
    \rho_{c}(r) \simeq\rho_{0}\left[1 + 0.091\cdot(r/r_{c})^{2}\right]^{-8}
\end{align}
where $r_{c}$ is the radius at which the density drops to one-half its peak value and the central density is given by
\begin{align}
\label{eq:3}
    \rho_{0}\simeq 3.1\times 10^{15}\left(\frac{2.5\times 10^{-22}\text{eV}}{m}\right)^{2}\left(\frac{\text{kpc}}{r_{c}}\right)^{4}\;\frac{M_{\odot}}{\text{Mpc}^{3}}.
\end{align}

As in \cite{Schive2014b}, we define the core mass $M_{c}$ as the mass enclosed by $r_{c}$ and note that in the case of a core with total mass $M$ it is
\begin{align}
\label{eq:4}
    \frac{M_{c}}{M_{\odot}} &\simeq 0.237\frac{M}{M_{\odot}}\\ 
    &\simeq 8.64\times 10^{6}\left(\frac{2.5\times 10^{-22}\text{eV}}{m}\right)^{2}\left(\frac{\text{kpc}}{r_{c}}\right)\nonumber\\ 
    &\simeq 1.81\times 10^{6}\left(\frac{2.5\times 10^{-22}\text{eV}}{m}\right)\left(\frac{E}{M}\right)^{1/2}\;\frac{\text{s}}{\text{km}}\, . \nonumber
\end{align}
\cite{Schive2014b} show evidence for the same scaling relation between $M_c, E,$ and $M$ for the final state of multiple core mergers where $E$ and $M$ refer to the total energy and mass of the system instead of just the core. We revisit this claim in \autoref{sec:multimerge} below.

The SP system and consequently the stationary solutions obey a scaling symmetry of the form \cite{Ji1994}:
\begin{align}
  \label{eq:5}
  \{t,x,U,\psi,\rho\}\rightarrow\{\lambda^{-2}\hat{t},\lambda^{-1}\hat{x},\lambda^{2}\hat{U},\lambda^{2}\hat{\psi},\lambda^{4}\hat{\rho}\},
\end{align}
where $\lambda$ is an arbitrary parameter. Note that $x\propto\rho^{-1/4}$ consistent with the relation between the average density of the core and its Jeans length \cite{Hu2000}. Throughout this paper we use an axion mass $m=2.5\times 10^{-22}$ eV.

\section{Numerical methods}
\label{sec:numerics}

The Schrödinger equation in comoving coordinates \cite{Woo2009} was implemented into the cosmological hydro code Nyx \cite{Almgren2013} in order to facilitate its later use for combined simulations of dark matter and baryons. The field $\psi$ is discretized on a grid as an additional dark matter component and integrated using a 4th order Runge-Kutta solver. We employed the multigrid Gauss-Seidel red-black Poisson solver provided by Nyx to compute the gravitational potential. The cosmological scale factor was set to $a = 1$ in all of the simulations reported here. All simulations used a grid size of $512^3$ cells.
 
In all runs, the total mass
\begin{align}
  \label{eq:6}
  M[\psi] = \int_{V}\rho\text{d}^{3}x
\end{align}
and energy
\begin{align}
  \label{eq:7}
  E[\psi] &= \int_{V}\left[\frac{\hbar^{2}}{2m^{2}}|\nabla\psi|^{2}+\frac{1}{2}U|\psi|^{2}\right]\text{d}^{3}x\\ \nonumber
          &= \int_{V}\frac{\hbar^{2}}{2m^{2}}(\nabla\sqrt{\rho})^{2}\text{d}^{3}x+ \int_{V}\frac{\rho}{2}v^{2}\text{d}^{3}x+ \int_{V}\frac{\rho}{2}U\text{d}^{3}x\\ \nonumber
          &= K_{\rho}+K_{v}+W
\end{align}
of the system was monitored. In the second line of \autoref{eq:7}, we used the Madelung representation \cite{Madelung1927}
\begin{align}
    \psi = \sqrt{\rho}e^{iS/\hbar}\quad,\quad v=\nabla S/m
\end{align}
and in the last line we divided the total energy into gradient energy  $K_{\rho}$, kinetic energy $K_{v}$, and potential energy $W$. Besides total energy, each contribution was measured separately in order to follow the dynamics of a particular system more closely. We use units $[M]=M_{\odot}$ and $[E]=M_{\odot}$km$^{2}$s$^{-2}$.

In addition, conservation of total angular momentum 
\begin{align}
\label{eq:8}
    L[\psi] &=\frac{1}{m}\int_{V}\psi^{*}\left[r\times(-i\hbar)\nabla\right]\psi\text{d}^{3}x\\ \nonumber
            &= \frac{1}{m}\int_{V}\left[r\times\rho \nabla S+\frac{i\hbar}{2} \nabla\times r\rho\right]\text{d}^{3}x\\\nonumber
            &=\int_{V}r\times\rho v\text{d}^{3}x
\end{align}
was verified assuming that the density falls off sufficiently rapidly that boundary terms vanish.

Owing to \autoref{eq:5}, the quantities defined above obey the scaling relations
\begin{align}
  \label{eq:9}
  \{M,K_{\rho},K_{v},W,L\rightarrow\lambda\hat{M},\lambda^{3}\hat{K_{\rho}},\lambda^{3}\hat{K_{v}},\lambda^{3}\hat{W},\lambda\hat{L}\}.
\end{align}

During the relaxation of the system, waves emitted by the merger carry mass and energy toward the numerical boundaries. In order to avoid spurious reheating from reflected waves, we follow
\cite{Guzman2004} and place a 'sponge' in the outer regions of the grid by adding an imaginary potential 
\begin{align}
    \label{eq:10}
    V(r) = &-\frac{i}{2}V_{0}\{2+\tanh[(r-r_{s})/\delta]-\tanh(r_{s}/\delta)\}\nonumber\\
    &\times\Theta[r-r_{p}]\,\,,
\end{align}
to the Schr\"odinger equation which efficiently absorbs matter. Here $r$ is the distance from the center of the numerical domain. The Heaviside function $\Theta$ ensures that the non-physical sponge is only added in the outer regions $r>r_{p}$. Let $r_{N}$ be half the box size. We then set $r_{p}=7/8 r_{N}$, $r_{s}=(r_{N}+r_{p})/2$, $\delta=(r_{N}-r_{p})$ and $V_{0}=0.6$. Although our numerical domains are always cubic, we use a spherical sponge since the final states of our simulations are approximately spherically symmetric.

In all runs, the time steps were chosen such that they fulfill the Courant-Friedrichs-Lewy (CFL) condition \cite{Ajaib2013}
\begin{align}
  \label{eq:11}
  \Delta t \leq \max\left[\frac{m}{6\hbar}\Delta x^{2},\frac{\hbar}{m|V|_{\text{max}}}\right]
\end{align}
where in all conducted runs the first argument is more stringent than the second. 

We tested our code by considering a single solitonic core. It was shown in \cite{Guzman2004,Bernal2006} that it is a virialized attractor solution of a broad class of initial conditions. Hence, we expect the core to be stable with low-amplitude excitations caused by numerical errors. 

The excitation manifests itself in a periodic variation in the central density. Its amplitude decreases faster than quadratically with resolution implying fast convergence of our code. The central density varies at most on the percent level if $r_c$  is resolved by at least 3 cells. For the simulations described below, the typical resolution is greater than 4 cells for all binary mergers and most multiple mergers. The oscillation frequency matches the one found in \cite{Guzman2004}. While kinetic and potential energy oscillate with opposite phase, total mass and energy are conserved to better than $10^{-3}$. The oscillation of $\psi$ in the complex plane has the expected frequency \cite{Guzman2004}.

We checked convergences of our code also for binary mergers. Increasing the resolution by a factor of two alters the results only negligibly. In all runs conserved quantities stay constant to better than $10^{-3}$ until matter is absorbed by the sponge.

\begin{figure*}
\includegraphics[width=0.325\textwidth]{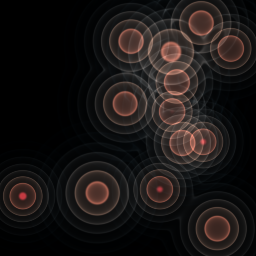}
\includegraphics[width=0.325\textwidth]{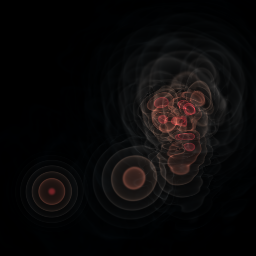}
\includegraphics[width=0.325\textwidth]{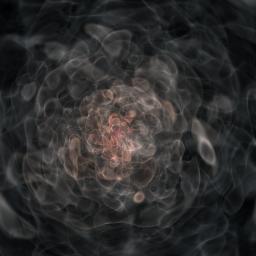}
\caption{Density distribution of a multimerger simulation with 13 halos at different times.}
\label{fig:13}
\end{figure*}

We use the yt toolkit \cite{Turk2011} for our analysis of numerical data and for the volume rendering of \autoref{fig:3b} and \autoref{fig:13}. Core profiles were fitted employing the radial density profile routines around the density maxima. Although cores with non-vanishing angular momentum are not expected to be perfectly spherical, we find that they can be well fitted by \autoref{eq:2}. Below, we therefore always assume spherical symmetry of the final state.

\section{Binary core mergers}
\label{sec:binary}

One of the distinctive features of hierarchical structure formation in FDM cosmologies is the presence of halo and subhalo cores evolving under a sequence of binary mergers which, to very good approximation, can be considered as isolated events. 

As a consequence of the scaling relations in \autoref{eq:5}, the initial conditions for an arbitrary binary collision are fully parametrized by few defining parameters, i.e. the relative velocity $v_{||}$ and distance $d$ between the cores, the mass ratio $\mu$ and total mass $M$, the phase difference $\Phi$, and the angular momentum $L_{z}$ perpendicular to the orbital plane chosen to be in the x-y-plane.

There are two distinct regimes. If the two cores are unbound ($E>0$) they superpose and pass through each other almost undisturbed \cite{Bernal2006,Gonzalez2011,Naraschewski1996,Roehrl1997}, behaving like solitons in this regime. If instead the cores are bound ($E<0$), they merge rapidly forming a new core \cite{Bernal2006}. Our main result is that the mass of the emerging core is largely independent of the initial angular momentum, distance and relative phase, but depends on the ratio of initial core masses and total energy.

\begin{figure*}
  \includegraphics[width=0.49\textwidth]{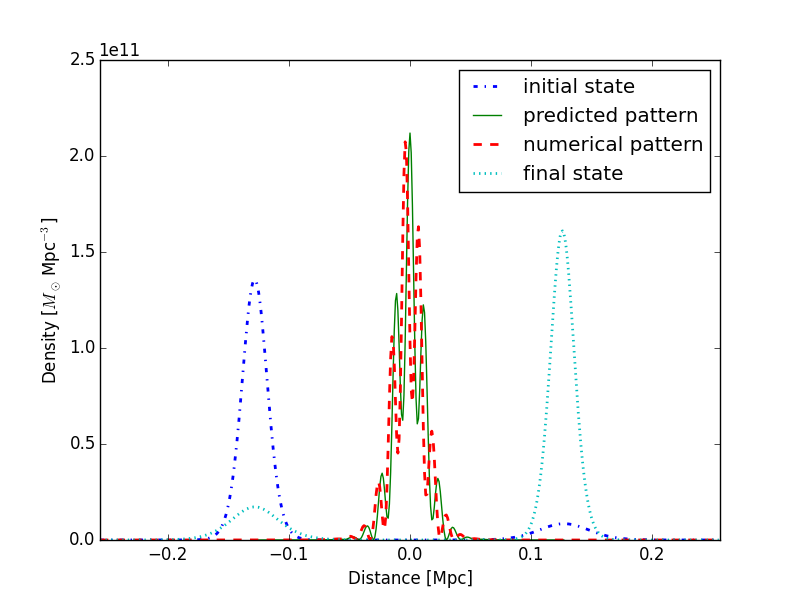}
  \hfill
  \includegraphics[width=0.49\textwidth]{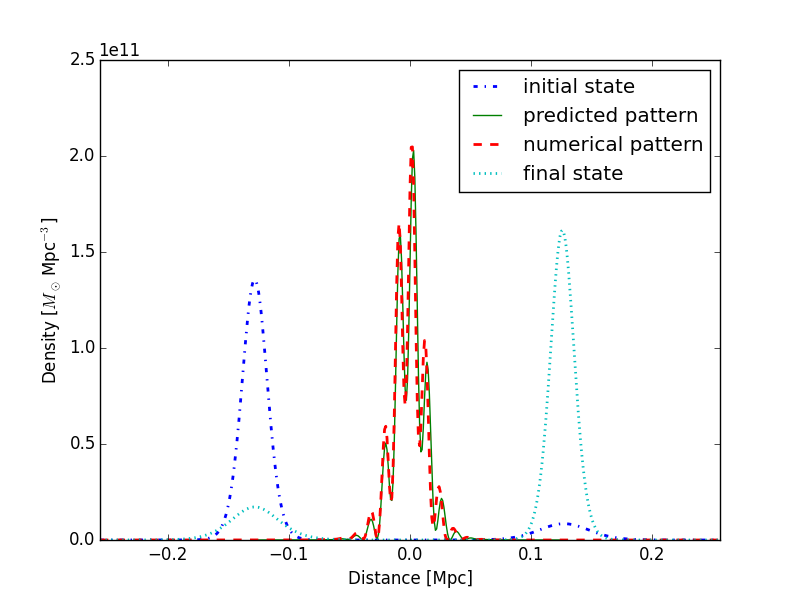}
  \includegraphics[width=0.49\textwidth]{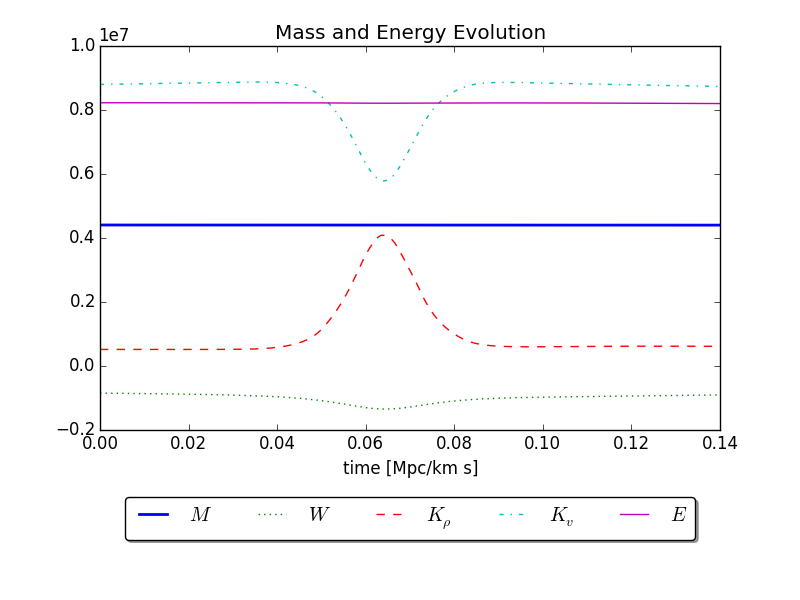}
  \hfill
  \includegraphics[width=0.49\textwidth]{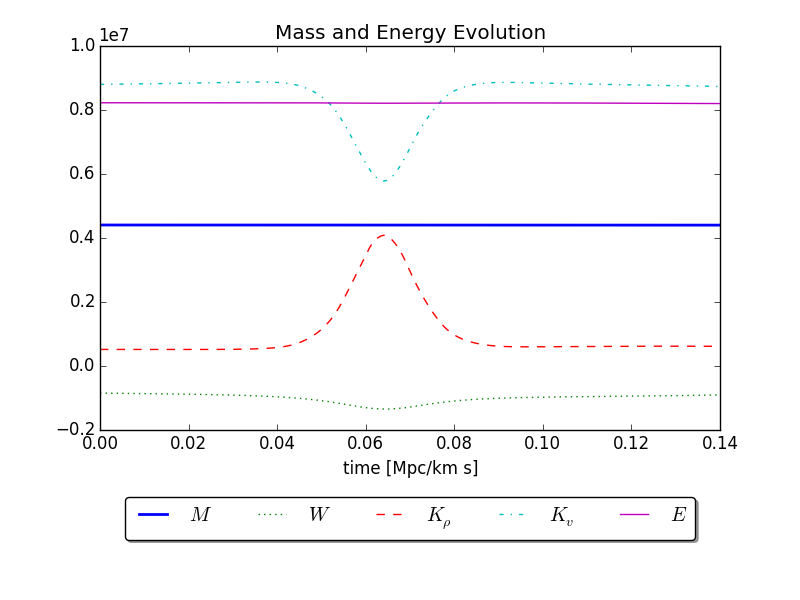}
  \caption{Head-on collision of two cores with mass ratio $\mu=2$ and high relative velocity. Upper panels: density profiles at different times for relative phases $\Phi=0$ (left) and $\Phi=\pi$ (right) along the symmetry axis. Numerical results are shown for the initial and final state as well as for the time of maximal interference. For comparison, we plot the interference pattern predicted from \autoref{eq:17} at the same time. Deviations can be attributed mostly to a small offset in the time of maximal interference. Lower panels: mass and energy contributions. Total energy and mass are conserved, while kinetic energy associated to the cores' relative motion ($K_{v}$) is transferred into the interference pattern yielding large values of $K_{\rho}$ during the interaction. The equality of the lower panels shows the independence of the evolution with respect to the inital phase shift $\Phi$.}
  \label{fig:2}
\end{figure*}

In order to analyse the unbound case, we consider two solitonic cores with $\mu\equiv M_{1}/M_{2}=M_{c,1}/M_{c,2}=2$, \mbox{$L_{z}=0\,M_{\odot}$ Mpc km/s} and $v_{||}=4\,$km/s. The cores are scaled such that the heavier one has a central density
\begin{align}
  \label{eq:12}
  \rho(0) = 1.36\times 10^{11}\, M_{\odot}\text{Mpc}^{-3},
\end{align}
roughly corresponding to the present cosmic critical density, giving a core radius $r_{c}\simeq 11.6\,\text{kpc}$. We emphasize that all results are independent of this overall scaling of the problem. The two cores are placed centrally in a $512$ kpc cubic box with $d=256\,$kpc yielding $E\simeq 8.2\times 10^{6}\,M_{\odot}$km$^{2}$s$^{-2}$.

\autoref{fig:2} shows the density profiles along the symmetry axis and the evolution of global quantities (mass and energy components) for two runs with relative phases $\Phi=0$ and $\Phi=\pi$. The final density distribution as well as the evolution of the global quantities are practically indistinguishable in both cases. Only the interference pattern at the time of superposition depends on the relative phase.

The observed interference pattern follows directly from a superposition of the two solitonic cores. Initially, the cores are placed at $\pm\hat{x}(t=0)=\pm d/2$. The corresponding wavefunction $\psi(t,x)$ is given by
\begin{align}
\label{eq:16}
  \psi(t,x) = &\, A_{1}(|x+\hat{x}|)e^{i(kx/2+\omega t+\Phi/2)}\\
  &+A_{2}(|x-\hat{x}|)e^{i(-kx/2+\omega t-\Phi/2)}\nonumber
\end{align}
where $(A_{1})^{2}$ and $(A_{2})^{2}$ are the density profiles of the two cores and $k=mv_{||}/\hbar$ is the wavenumber corresponding to their relative velocity. The time $t_{\text{int}}$ of maximal interference is defined by $\hat{x}(t_{\text{int}})=0$. At that time,
\begin{align}
\label{eq:17}
  |\psi(t_{\text{int}},x)|^{2} =&\,A_{1}(|x|)^{2} + A_{2}(|x|)^{2}\\
  &+ 2A_{1}(|x|)A_{2}(|x|)\cos(kx+\Phi))\,\,.\nonumber
\end{align}
We thus expect that the period of the interference pattern is given by the de Broglie wave length 
\begin{align}
\label{eq:18}
  \lambda = \frac{2\pi}{k} = \frac{2\pi\hbar}{mv_{||}}
\end{align}
corresponding to the relative velocity. Here, $\lambda\simeq 12$ kpc. It is therefore well resolved by 12 cells.

\begin{figure*}
  \includegraphics[width=0.49\textwidth]{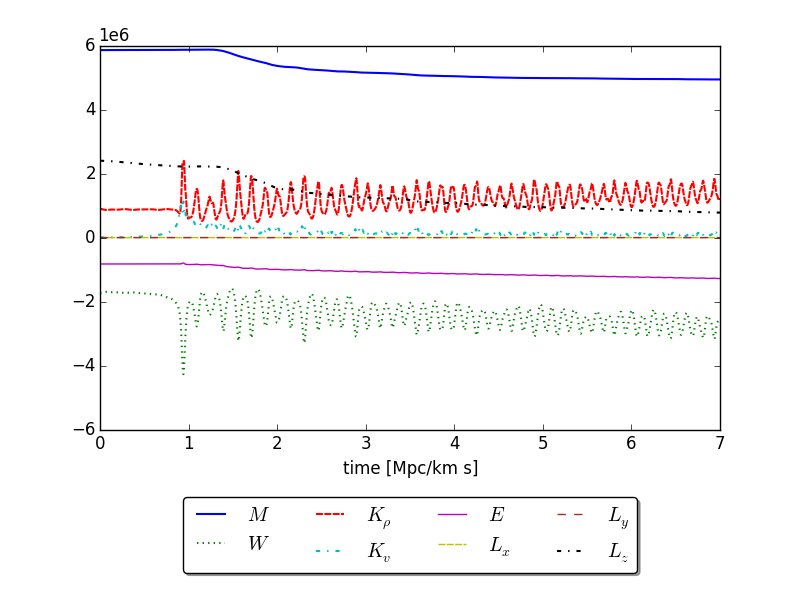}
  \hfill
  \includegraphics[width=0.49\textwidth]{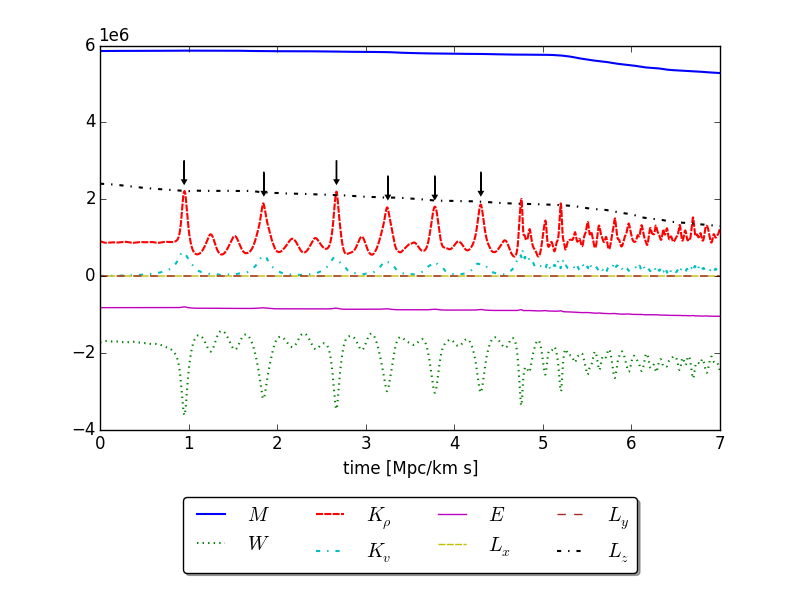}
  \caption{Mass, energy and angular momentum evolution of two representative binary collisions with initial values $\mu=1$, $v_{||}=0$ km/s, and $L_{z}=2.4\times 10^{4}\,M_{\odot}$ Mpc km/s (rescaled by $10^2$). Cores with equal phase ($\Phi=0$) immediately merge (left). In perfect phase opposition ($\Phi=\pi$), the two cores first mutually repel each other multiple times before merging (right). The bounces are indicated by black arrows. The emerging cores are excited as seen by the oscillations of gradient and gravitational energy, $K_{\rho}$ and $W$. The loss of total mass, energy, and angular momentum results from matter absorption inside the sponge.}
  \label{fig:3}
\end{figure*}

The interference pattern predicted by \autoref{eq:17} matches the numerical results as seen in \autoref{fig:2}. During the interaction, gravity slightly contracts the density profiles. Neglecting this small effect, we see that they remain in a superposition state of two solitonic cores even during their interaction. As expected, the potential energy mildly increases during the collision, while mass and total energy are conserved. During the collision, the kinetic energy from the cores' relative motion is stored in the interference pattern, strongly boosting the gradient energy contribution $K_{\rho}$. At later times, the energy is transferred back to the cores' motion. There is no significant decrease in velocity or deformation of the density profiles due to the collision. The cores thus indeed behave like solitons in this regime.

The evolution of a bound binary system with negative total energy is very different. In this case, the cores rapidly merge and relax to a new solitonic core by gravitational cooling \cite{Bernal2006b}. One interesting exception is the case of binary collisions with perfect phase opposition $\Phi=\pi$ and equal masses $\mu=1$ during which the destructive interference gives rise to a repulsive effect, causing the cores to bounce off each other \cite{Paredes2016}.

\begin{figure}
\includegraphics[width=\columnwidth]{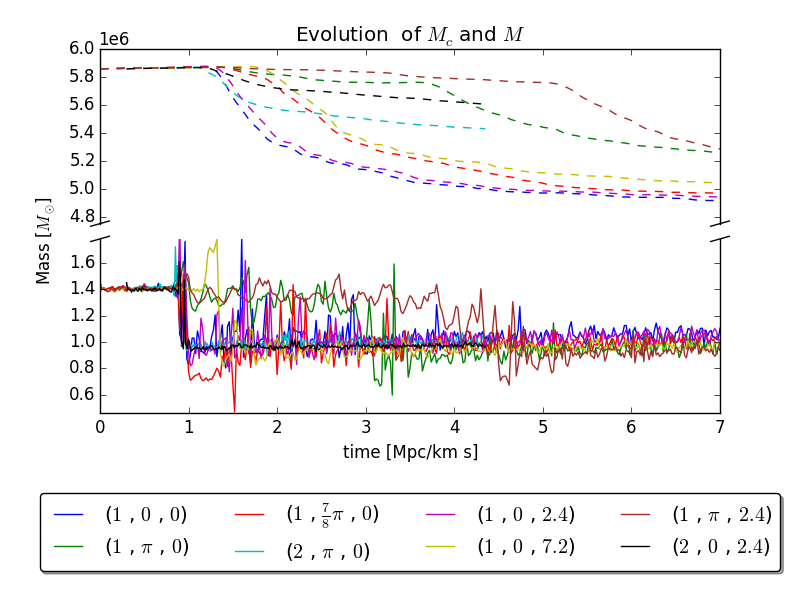}
\caption{Evolution of the core (solid lines) and total (dashed lines) mass for binary mergers. The triplets identify the point $(\mu,\Phi,L_{z})$ in parameter space. Angular momentum is given in units of $\left[L_{z}\right] = 10^{4}\,M_{\odot}$ Mpc km/s.}
\label{fig:8}
\end{figure}

For our study of bound binary collisions, we placed two halos along the central axis in a $1024$ kpc cubic box with $d=256\,$kpc. As before, the cores are scaled such that the central density of the heavier core obeys $\rho(0)=\rho_{\text{cr}}$. We need the larger box compared to the previous runs since the two halos emit mass while merging. We require this mass to be able to propagate sufficiently far away from the merger before being absorbed inside the sponge.

\begin{figure*}
\includegraphics[width=0.24\textwidth]{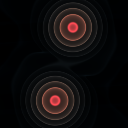}
\includegraphics[width=0.24\textwidth]{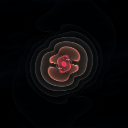}
\includegraphics[width=0.24\textwidth]{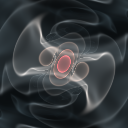}
\includegraphics[width=0.24\textwidth]{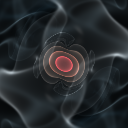}
\includegraphics[width=0.24\textwidth]{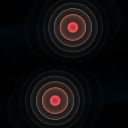}
\includegraphics[width=0.24\textwidth]{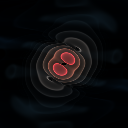}
\includegraphics[width=0.24\textwidth]{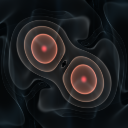}
\includegraphics[width=0.24\textwidth]{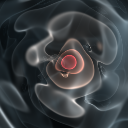}
\caption{Volume rendered images of two representative binary mergers in phase (top) and with opposite phase (bottom) showing the central region of the computational domain at $t= 0.7$, $t=0.94$, $t=2.0$ and $t=7.0$ in Mpc/km s.}
\label{fig:3b}
\end{figure*}

In \autoref{fig:3}, we show the mass, energy and angular momentum evolution of two representative runs with $\mu=1$, $v_{||}=0$ km/s and $L_{z}=2.4\times 10^{4}\,M_{\odot}$ Mpc km/s. We again emphasize that the system can be arbitrarily rescaled using \autoref{eq:5} without changing the results. On the left, the two cores are in phase. 
They merge after approximately one free fall time, $t_{\text{ff}}\simeq 0.94$ Mpc/km s, and form a new excited solitonic core within roughly one oscillation period. The core's frequency $f\simeq 8\,$km/Mpc/s, implies that it consists of only $70\%$ of the initial mass \cite{Guzman2004} whereas approximately $30\,$\% of the initial total mass was radiated off by gravitational cooling. This estimate is confirmed by the evolution of the total core mass $M_{c}=M_{c,1}+M_{c,2}$ and the total mass $M$ shown in \autoref{fig:8}. Initially, $M_{c}\simeq\frac{1}{4}M$ as expected, decreasing roughly by $30$\% during the merger. After a while, the ejected mass reaches the sponge and is absorbed. This does not alter the results, since in all conducted runs, the ejected mass is roughly an order of magnitude above the escape velocity $v_{\text{esc}}=\sqrt{2GM/r}$ and will not fall back onto the core.

\begin{figure}
\includegraphics[width=\columnwidth]{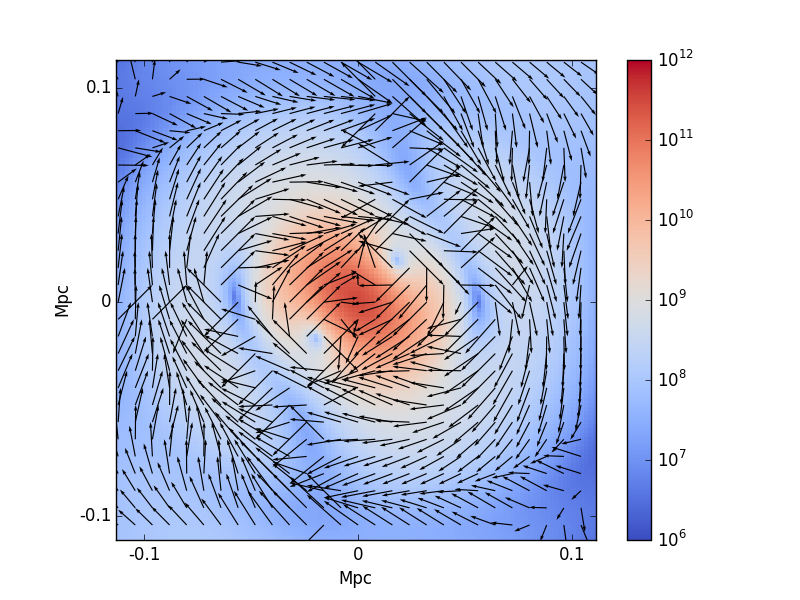}
\caption{Slice through the symmetry plane of a representative ellipsoid. Its density is color-coded while arrows denote the strength and direction of its velocity field. It roughly forms closed elliptical orbits.}
\label{fig:3c}
\end{figure}

In the case of solitonic cores with equal mass ($\mu = 1$) but opposite phase ($\Phi=\pi$), the destructive interference gives rise to a repulsive interaction, causing the cores to bounce off each other several times before merging (cf. right panel of \autoref{fig:3}). This behaviour was also observed in \cite{Paredes2016}. The arrows indicate the bounces which result in a noticeable compression of the individual cores. Radiation produced by each encouter results in a damping of the bounces and a decreasing amplitude of the compression. Eventually, the symmetry is broken by the accumulation of small numerical errors producing a slight phase shift, causing the cores to merge in the end. At later times, the evolution is qualitatively identical to the case with $\Phi=0$ as can be seen by comparing the core and halo mass evolution in \autoref{fig:8}.

Volume rendered images of both runs are shown in \autoref{fig:3b}. Especially in the upper panels, a noticeable eccentricity of the newly formed core can be recognized. These rotating ellipsoids are qualitatively those investigated in \cite{RindlerDaller2009,RindlerDaller2011,RindlerDaller2012,RindlerDaller2013a,RindlerDaller2013b}. In particular, their internal velocity fields roughly confine density distributions on elliptical orbits. A slice through a representative ellipsoid is shown in \autoref{fig:3c}. Further analysis will be the subject of future work.

We tested the sensitivity of the repulsive interaction to small deviations from exact phase opposition by considering a phase difference $\Phi=7/8\pi$. In this case, only a single bounce occurs before the cores merge. Similarly, for a mass ratio $\mu=2$ and $\Phi=\pi$ the cores merge without any observable repulsion. These results suggest that in any realistic scenario absent finely tuned phase opposition and mass equality, repulsive behavior of colliding solitonic cores can be ignored for all practical purposes.

\begin{figure*}
\includegraphics[width=0.49\textwidth]{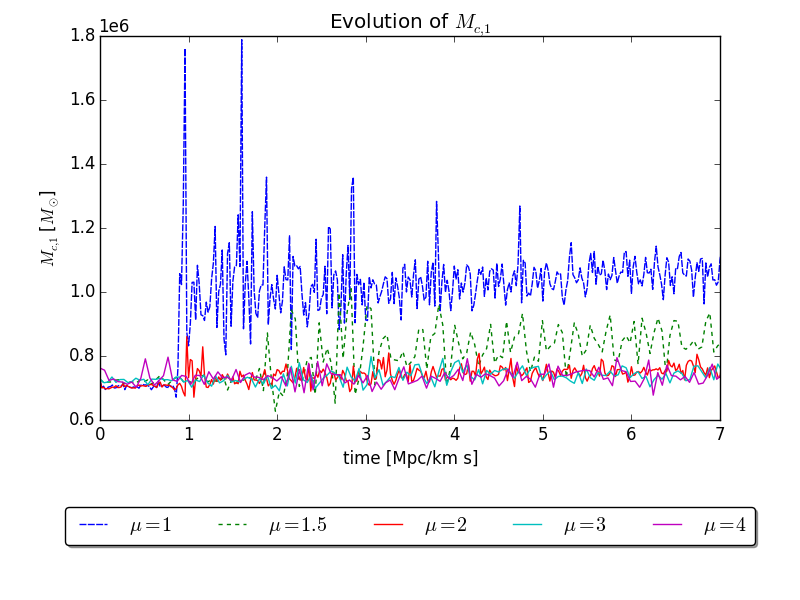}
\hfill
\includegraphics[width=0.49\textwidth]{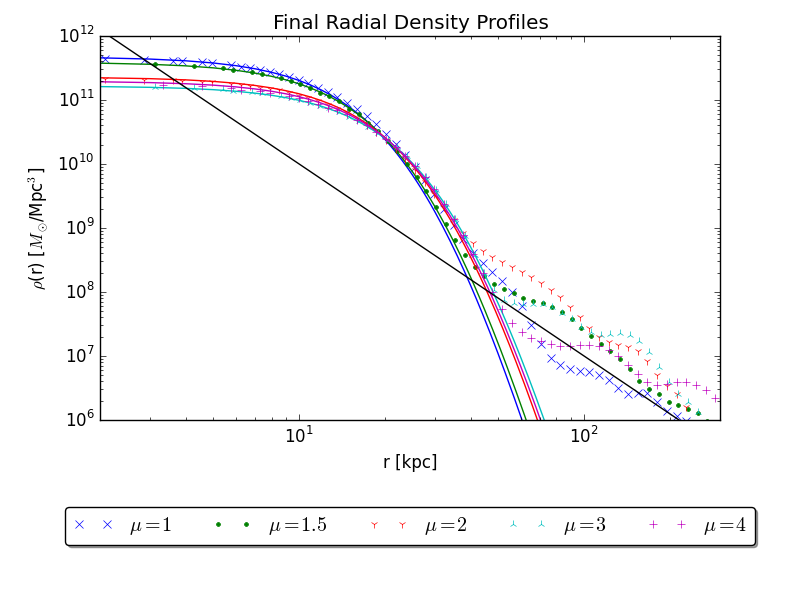}
\caption{Binary mergers with different mass ratios $\mu$. Left: evolution of the core mass of the more massive core. Right: final radial density profiles. Solid lines represent fitted core profiles as defined in \autoref{eq:2}. The black line corresponds to $r^{-3}$ as expected for the outer parts of an NFW profile.}
\label{fig:9}
\end{figure*}

\begin{figure}
\includegraphics[width=\columnwidth]{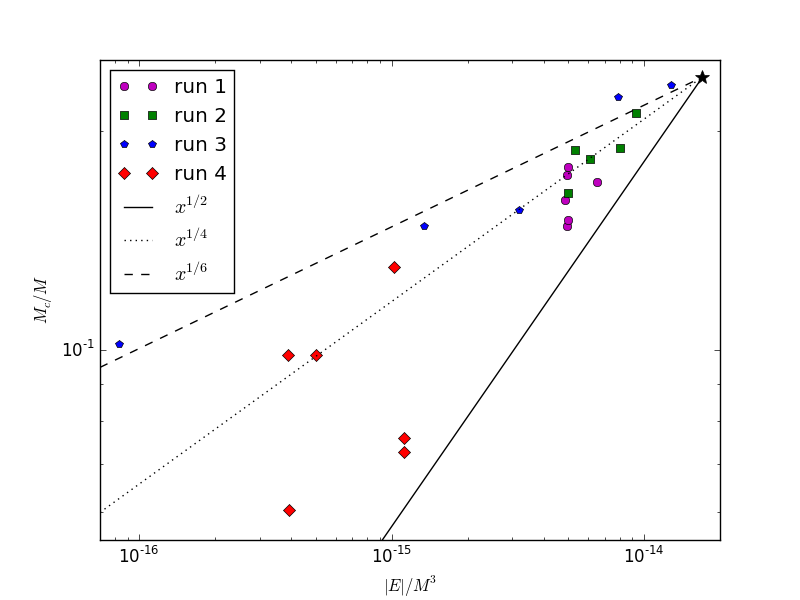}
\caption{Core mass as a function of the total energy and mass. The star indicates the relation for a single solitonic core. Run 1 denotes the simulations with almost equal total energy for different angular momenta and phases. Runs 2 and 3 show the dependence on mass ratio $\mu$ and total energy $E$, respectively. Multiple core mergers are shown as run 4 (cf. \autoref{sec:multimerge}). See main text for details.}
\label{fig:10}
\end{figure}

We conducted a series of binary mergers spanning the parameter space $(\mu,\Phi,L_{z})$. For all runs, we set \mbox{$v_{||}=0$ km/s}, $\mu \le 2$, and  $L_{z} \le 7.2 \times 10^{4}\,M_{\odot}$ Mpc km/s so that the cores are bounded and overlap when reaching the semi-minor axis. Our main result is that the core mass evolution is nearly independent of these parameters within the considered ranges. In all cases, the mass of the emerging core is approximately $70\%$ of the sum of the progenitors' core masses. The core and total mass evolution of eight representative runs are shown in \autoref{fig:8}. The ratio between final core and total masses is approximately one fifth implying that $80\%$ of the remaining bound mass resides in the solitonic core while the remainder has formed a diffuse halo around it. Note that due to the restriction to small angular momenta and mass ratios, the total energy varies only very little for all runs. The energy dependent final core masses $M_{c}(E)$ of the above runs are shown in \autoref{fig:10} (run 1).  

Assuming a constant fraction of final to initial core masses of $\sim 70\%$ even for $\mu \ne 1$ implies that the final core is less massive than the more massive progenitor if $\mu \gtrsim 7/3$. We therefore expect the change of $M_c$ of the more massive core to saturate at roughly this mass ratio. This is qualitatively confirmed by our simulations. For $\mu \gtrsim 2$, the less massive core is completely disrupted and forms a diffuse halo. \autoref{fig:9} shows the core mass evolution for different mass ratios (left). Here, the initial core mass corresponds to the more massive core. On the right, the final radial density profiles can be seen. They consist of a solitonic core well fitted by \autoref{eq:2} and a shallow outer tail. Interestingly, the tails in all cases approximately follow a power law decline with a logarithmic slope of roughly $-3$ as expected for the outer parts of a Navarro-Frenk-White (NFW) halo profile. This behavior is consistent with the results of \cite{Schive2014b} but finding NFW-like halos already in the case of binary mergers suggests that it may be more robust than previously expected.

The fitted core masses are mildly energy dependent as can be seen in \autoref{fig:10} (run 2). They very broadly follow a power law with
\begin{align}
\label{eq:20}
\frac{M_{c}}{M} = 656\left(\frac{|E|}{M^{3}}\right)^{1/4}M_{\odot}^{1/2}\text{km}^{-1/2}\text{s}^{1/2}.
\end{align}
In \autoref{fig:10}, the final core mass $M_{c}$ is normalized to the initial total mass $M$ in order to obtain an invariant relation with respect to the scaling properties given in \autoref{eq:9}. For a single solitonic core, $M_{c}/M\simeq 0.237$ and $|E|/M^{3}\simeq 1.7\times 10^{-14}M_{\odot}^{-2}\text{km}^{2}\text{s}^{-2}$ as indicated by the black star in the upper right corner. This point is consistent with \autoref{eq:20} since a single core is the limit of infinite mass ratio. A single core is the ground state solution of the SP system. It is therefore the point of minimum energy and maximum core mass per total mass.

Finally, we conducted a series of runs with $\Phi=0$, $\mu=1$, $L_{z}=0$ and varying $d$ and $v_{||}$ over a wide range of energies. The fitted final core masses are collectively shown in \autoref{fig:10} (run 3). The dashed line corresponds to 
\begin{align}
\label{eq:21}
\frac{M_{c}}{M} = 46.7\left(\frac{|E|}{M^{3}}\right)^{1/6}M_{\odot}^{1/3}\text{km}^{-1/3}\text{s}^{1/3}\,\,,
\end{align}
indicating a weaker energy dependence for $\mu=1$ than for larger mass ratios.

In conclusion, our results for binary mergers show consistently that the final core mass does not depend on initial phase difference but only on mass ratio, total initial mass, and total energy of the system. It depends on angular momentum, relative distance and velocity only via the total energy.

\begin{figure}
\includegraphics[width=\columnwidth]{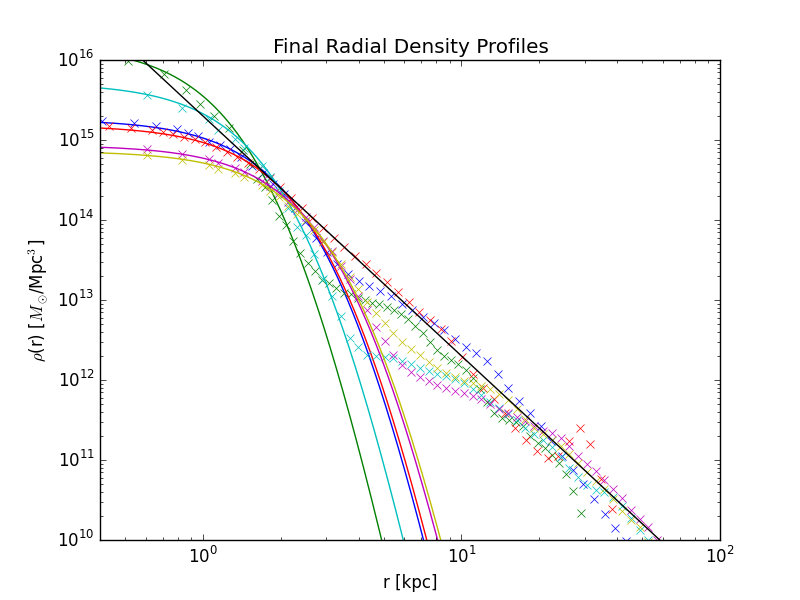}
\caption{Final radial density profiles for all conducted multimerger runs. Solid lines represent fitted core profiles as defined in \autoref{eq:2}. The black line corresponds to $r^{-3}$ as expected for the outer parts of an NFW profile.}
\label{fig:12}
\end{figure}

\section{Mergers of Multiple Cores}
\label{sec:multimerge}

In order to study more complex, non-equlibrium problems we follow \cite{Schive2014b} and investigate mergers of multiple cores. From our previous analysis we know that the merging time of binaries is negligible with respect to the typical free-fall time. We can therefore safely assume that a multimerger consists of a series of binary mergers within a deeper gravitational well.

For all runs, we draw halo masses from a Gaussian distribution within the $2\sigma$-band around a chosen average halo mass. We then place the halos uniformly inside the central numerical domain, rejecting positions that would result in an overlap of halos or close proximity to the outer sponge. Rejected halo positions are redrawn until acceptable. Halos are initialized with random phases. We simulated multimergers of up to 13 halos. As a typical example, \autoref{fig:13} shows the volume rendered images of a multimerger with 13 halos at three different times.

The final radial density profiles for all runs are presented in \autoref{fig:12}. As in the case of binary mergers and in full agreement with \cite{Schive2014b}, their central regions can be fitted with a solitonic core profile, \autoref{eq:2}, while the tails fall off like $r^{-3}$ consistent with the outer profile of an NFW halo. The final core masses are summarized in \autoref{fig:10} (run 4). We cannot confirm the $M_{c}\sim (E/M)^{1/2}$ scaling shown in \cite{Schive2014b} which may in part be a consequence of the fact that, in contrast with their analysis, all results in \autoref{fig:12} are normalized to the initial total mass $M$. This eliminates any scaling with energy originating only from the scale invariance of the SP system, making the results more sensitive to the intrinsic energy dependence of multimergers. We verified that this discrepancy is unrelated to the initial phase shifts of individual halos. 

\section{Conclusions}
\label{sec:conclusion}

We presented an investigation of merging solitonic halo cores in full three-dimensional simulations of the Schrödinger-Poisson (SP) equations without assuming any symmetries. These cores have been predicted to form in the center of ultra-light axion dark matter halos. Their structure is identical to Newtonian oscillaton solutions also known as boson stars. 

Our results demonstrate a number of robust features of binary core mergers. Qualitatively, bound systems rapidly merge within roughly one oscillation period of the emerging core after approaching to a distance at which the characteristic core radii overlap. It was shown in \cite{Guzman2016} that luminous matter cannot follow these extreme dynamics and is expelled from the gravitational potential.

During this dynamical phase, gravitational cooling is most efficient and essentially determines the loss of mass and angular momentum of the merged core, while continuing to dampen its excitations during the ensuing several oscillation periods. One exception is the case of perfect phase opposition and equal masses in which case the cores initially repel each other, leading to a bouncing behavior until small accumulated phase differences again cause a rapid merger on a dynamical time scale. Owing to the fine tuning required for this situation, we do not consider it relevant in the context of cosmology.

The mass of the emerging core does not directly depend on the binary angular momentum, initial distance, and phase shift between the solitonic cores. It does depend weakly on their mass ratio and total energy. The mass of the more massive core can only be enhanced by binary mergers with mass ratio $\mu<7/3$. Otherwise, the smaller core is completely disrupted and forms an NFW-like halo around the more massive one.

Neither for the binary mergers nor for the sample of multiple core mergers we were able to reproduce the scaling of core mass with total energy and mass, $M_{c}\sim (E/M)^{1/2}$, reported in \cite{Schive2014b}. After normalizing our results to equal total mass using the scale invariance of the SP equations in order to eliminate spurious scaling behavior, we find no convincing evidence for a universal scaling of core mass with total energy. More detailed analysis with larger ranges of $M_c$ and $E$ and comparison to cosmological simulations of the SP system are needed to further elucidate this discrepancy. 

The final states of both the binary and multimergers are roughly spherical symmetric. We confirm that their radial density profiles consist of a solitonic core well modeled by \autoref{eq:2} and an NFW-like outer region falling off as $r^{-3}$ \cite{Schive2014b}. If the system is initialized with non-zero total angular momentum, we qualitatively recover the rotating ellipsoidal cores studied in \cite{RindlerDaller2009,RindlerDaller2011,RindlerDaller2012,RindlerDaller2013a,RindlerDaller2013b}.

Our results are useful for a refined modeling of the properties of halo cores in FDM cosmologies, for instance in stochastic merger tree realizations of the halo and subhalo population \cite{Du2016}. This approach complements other simplified structure formation models that include the effects of the linear transfer function \cite{Schive2016,Sarkar2016} and mass-dependent collapse barrier \cite{Marsh2013,Bozek2015,Marsh2016} but neglect the presence of solitonic cores. They might also help to understand the relation of core and halo masses in cosmological FDM simulations \cite{Schive2014b}. Eventually, the consequences of solitonic cores for galaxy evolution will have to be better understood in order to tighten the constraints on ultra-light axion masses from reionization, the UV luminosity function, or halo substructure. Simulations of more realistic cosmological setups including baryonic physics are in preparation.

\acknowledgements
We thank C. Behrens, X. Du, D.J.E. Marsh, and J. Veltmaat for helpful discussions. The simulations were performed with resources provided by the North-German Supercomputing Alliance (HLRN).

%\bibliography{multimergers}

\begin{thebibliography}{49}%
\makeatletter
\providecommand \@ifxundefined [1]{%
 \@ifx{#1\undefined}
}%
\providecommand \@ifnum [1]{%
 \ifnum #1\expandafter \@firstoftwo
 \else \expandafter \@secondoftwo
 \fi
}%
\providecommand \@ifx [1]{%
 \ifx #1\expandafter \@firstoftwo
 \else \expandafter \@secondoftwo
 \fi
}%
\providecommand \natexlab [1]{#1}%
\providecommand \enquote  [1]{``#1''}%
\providecommand \bibnamefont  [1]{#1}%
\providecommand \bibfnamefont [1]{#1}%
\providecommand \citenamefont [1]{#1}%
\providecommand \href@noop [0]{\@secondoftwo}%
\providecommand \href [0]{\begingroup \@sanitize@url \@href}%
\providecommand \@href[1]{\@@startlink{#1}\@@href}%
\providecommand \@@href[1]{\endgroup#1\@@endlink}%
\providecommand \@sanitize@url [0]{\catcode `\\12\catcode `\$12\catcode
  `\&12\catcode `\#12\catcode `\^12\catcode `\_12\catcode `\%12\relax}%
\providecommand \@@startlink[1]{}%
\providecommand \@@endlink[0]{}%
\providecommand \url  [0]{\begingroup\@sanitize@url \@url }%
\providecommand \@url [1]{\endgroup\@href {#1}{\urlprefix }}%
\providecommand \urlprefix  [0]{URL }%
\providecommand \Eprint [0]{\href }%
\providecommand \doibase [0]{http://dx.doi.org/}%
\providecommand \selectlanguage [0]{\@gobble}%
\providecommand \bibinfo  [0]{\@secondoftwo}%
\providecommand \bibfield  [0]{\@secondoftwo}%
\providecommand \translation [1]{[#1]}%
\providecommand \BibitemOpen [0]{}%
\providecommand \bibitemStop [0]{}%
\providecommand \bibitemNoStop [0]{.\EOS\space}%
\providecommand \EOS [0]{\spacefactor3000\relax}%
\providecommand \BibitemShut  [1]{\csname bibitem#1\endcsname}%
\let\auto@bib@innerbib\@empty
%</preamble>
\bibitem [{\citenamefont {Marsh}\ and\ \citenamefont
  {Ferreira}(2010)}]{Marsh2010}%
  \BibitemOpen
  \bibfield  {author} {\bibinfo {author} {\bibfnamefont {D.~J.}\ \bibnamefont
  {Marsh}}\ and\ \bibinfo {author} {\bibfnamefont {P.~G.}\ \bibnamefont
  {Ferreira}},\ }\href {\doibase 10.1103/PhysRevD.82.103528} {\bibfield
  {journal} {\bibinfo  {journal} {Phys.Rev.}\ }\textbf {\bibinfo {volume}
  {D82}},\ \bibinfo {pages} {103528} (\bibinfo {year} {2010})},\ \Eprint
  {http://arxiv.org/abs/1009.3501} {arXiv:1009.3501 [hep-ph]} \BibitemShut
  {NoStop}%
%%CITATION = ARXIV:1009.3501;%%
\bibitem [{\citenamefont {Marsh}\ and\ \citenamefont {Silk}(2013)}]{Marsh2013}%
  \BibitemOpen
  \bibfield  {author} {\bibinfo {author} {\bibfnamefont {D.~J.~E.}\
  \bibnamefont {Marsh}}\ and\ \bibinfo {author} {\bibfnamefont
  {J.}~\bibnamefont {Silk}},\ }\href {\doibase 10.1093/mnras/stt2079}
  {\bibfield  {journal} {\bibinfo  {journal} {Mon.Not.Roy.Astron.Soc.}\
  }\textbf {\bibinfo {volume} {437}},\ \bibinfo {pages} {2652} (\bibinfo {year}
  {2013})},\ \Eprint {http://arxiv.org/abs/1307.1705} {arXiv:1307.1705
  [astro-ph.CO]} \BibitemShut {NoStop}%
%%CITATION = ARXIV:1307.1705;%%
\bibitem [{\citenamefont {{Bozek}}\ \emph {et~al.}(2015)\citenamefont
  {{Bozek}}, \citenamefont {{Marsh}}, \citenamefont {{Silk}},\ and\
  \citenamefont {{Wyse}}}]{Bozek2015}%
  \BibitemOpen
  \bibfield  {author} {\bibinfo {author} {\bibfnamefont {B.}~\bibnamefont
  {{Bozek}}}, \bibinfo {author} {\bibfnamefont {D.~J.~E.}\ \bibnamefont
  {{Marsh}}}, \bibinfo {author} {\bibfnamefont {J.}~\bibnamefont {{Silk}}}, \
  and\ \bibinfo {author} {\bibfnamefont {R.~F.~G.}\ \bibnamefont {{Wyse}}},\
  }\href {\doibase 10.1093/mnras/stv624} {\bibfield  {journal} {\bibinfo
  {journal} {Mon. Not. Roy. Astron. Soc.}\ }\textbf {\bibinfo {volume} {450}},\
  \bibinfo {pages} {209} (\bibinfo {year} {2015})},\ \Eprint
  {http://arxiv.org/abs/1409.3544} {arXiv:1409.3544} \BibitemShut {NoStop}%
\bibitem [{\citenamefont {{Marsh}}\ and\ \citenamefont
  {{Pop}}(2015)}]{Marsh2015b}%
  \BibitemOpen
  \bibfield  {author} {\bibinfo {author} {\bibfnamefont {D.~J.~E.}\
  \bibnamefont {{Marsh}}}\ and\ \bibinfo {author} {\bibfnamefont {A.-R.}\
  \bibnamefont {{Pop}}},\ }\href {\doibase 10.1093/mnras/stv1050} {\bibfield
  {journal} {\bibinfo  {journal} {Mon. Not. Roy. Astron. Soc.}\ }\textbf
  {\bibinfo {volume} {451}},\ \bibinfo {pages} {2479} (\bibinfo {year}
  {2015})},\ \Eprint {http://arxiv.org/abs/1502.03456} {arXiv:1502.03456}
  \BibitemShut {NoStop}%
\bibitem [{\citenamefont {Arvanitaki}\ \emph {et~al.}(2010)\citenamefont
  {Arvanitaki}, \citenamefont {Dimopoulos}, \citenamefont {Dubovsky},
  \citenamefont {Kaloper},\ and\ \citenamefont
  {March-Russell}}]{Arvanitaki2010}%
  \BibitemOpen
  \bibfield  {author} {\bibinfo {author} {\bibfnamefont {A.}~\bibnamefont
  {Arvanitaki}}, \bibinfo {author} {\bibfnamefont {S.}~\bibnamefont
  {Dimopoulos}}, \bibinfo {author} {\bibfnamefont {S.}~\bibnamefont
  {Dubovsky}}, \bibinfo {author} {\bibfnamefont {N.}~\bibnamefont {Kaloper}}, \
  and\ \bibinfo {author} {\bibfnamefont {J.}~\bibnamefont {March-Russell}},\
  }\href {\doibase 10.1103/PhysRevD.81.123530} {\bibfield  {journal} {\bibinfo
  {journal} {Phys.Rev.}\ }\textbf {\bibinfo {volume} {D81}},\ \bibinfo {pages}
  {123530} (\bibinfo {year} {2010})},\ \Eprint {http://arxiv.org/abs/0905.4720}
  {arXiv:0905.4720 [hep-th]} \BibitemShut {NoStop}%
%%CITATION = ARXIV:0905.4720;%%
\bibitem [{\citenamefont {Hu}\ \emph {et~al.}(2000)\citenamefont {Hu},
  \citenamefont {Barkana},\ and\ \citenamefont {Gruzinov}}]{Hu2000}%
  \BibitemOpen
  \bibfield  {author} {\bibinfo {author} {\bibfnamefont {W.}~\bibnamefont
  {Hu}}, \bibinfo {author} {\bibfnamefont {R.}~\bibnamefont {Barkana}}, \ and\
  \bibinfo {author} {\bibfnamefont {A.}~\bibnamefont {Gruzinov}},\ }\href
  {\doibase 10.1103/PhysRevLett.85.1158} {\bibfield  {journal} {\bibinfo
  {journal} {Phys.Rev.Lett.}\ }\textbf {\bibinfo {volume} {85}},\ \bibinfo
  {pages} {1158} (\bibinfo {year} {2000})},\ \Eprint
  {http://arxiv.org/abs/astro-ph/0003365} {arXiv:astro-ph/0003365 [astro-ph]}
  \BibitemShut {NoStop}%
%%CITATION = ASTRO-PH/0003365;%%
\bibitem [{\citenamefont {Woo}\ and\ \citenamefont {Chiueh}(2009)}]{Woo2009}%
  \BibitemOpen
  \bibfield  {author} {\bibinfo {author} {\bibfnamefont {T.-P.}\ \bibnamefont
  {Woo}}\ and\ \bibinfo {author} {\bibfnamefont {T.}~\bibnamefont {Chiueh}},\
  }\href {\doibase 10.1088/0004-637X/697/1/850} {\bibfield  {journal} {\bibinfo
   {journal} {Astrophys.J.}\ }\textbf {\bibinfo {volume} {697}},\ \bibinfo
  {pages} {850} (\bibinfo {year} {2009})},\ \Eprint
  {http://arxiv.org/abs/0806.0232} {arXiv:0806.0232 [astro-ph]} \BibitemShut
  {NoStop}%
%%CITATION = ARXIV:0806.0232;%%
\bibitem [{\citenamefont {{Chavanis}}(2011)}]{Chavanis2011}%
  \BibitemOpen
  \bibfield  {author} {\bibinfo {author} {\bibfnamefont {P.-H.}\ \bibnamefont
  {{Chavanis}}},\ }\href {\doibase 10.1103/PhysRevD.84.043531} {\bibfield
  {journal} {\bibinfo  {journal} {prd}\ }\textbf {\bibinfo {volume} {84}},\
  \bibinfo {eid} {043531} (\bibinfo {year} {2011})},\ \Eprint
  {http://arxiv.org/abs/1103.2050} {arXiv:1103.2050} \BibitemShut {NoStop}%
\bibitem [{\citenamefont {{Su{\'a}rez}}\ and\ \citenamefont
  {{Matos}}(2011)}]{Matos2011}%
  \BibitemOpen
  \bibfield  {author} {\bibinfo {author} {\bibfnamefont {A.}~\bibnamefont
  {{Su{\'a}rez}}}\ and\ \bibinfo {author} {\bibfnamefont {T.}~\bibnamefont
  {{Matos}}},\ }\href {\doibase 10.1111/j.1365-2966.2011.19012.x} {\bibfield
  {journal} {\bibinfo  {journal} {mnras}\ }\textbf {\bibinfo {volume} {416}},\
  \bibinfo {pages} {87} (\bibinfo {year} {2011})},\ \Eprint
  {http://arxiv.org/abs/1101.4039} {arXiv:1101.4039 [gr-qc]} \BibitemShut
  {NoStop}%
\bibitem [{\citenamefont {{Martinez-Medina}}\ and\ \citenamefont
  {{Matos}}(2014)}]{Matos2014}%
  \BibitemOpen
  \bibfield  {author} {\bibinfo {author} {\bibfnamefont {L.~A.}\ \bibnamefont
  {{Martinez-Medina}}}\ and\ \bibinfo {author} {\bibfnamefont {T.}~\bibnamefont
  {{Matos}}},\ }\href {\doibase 10.1093/mnras/stu1453} {\bibfield  {journal}
  {\bibinfo  {journal} {mnras}\ }\textbf {\bibinfo {volume} {444}},\ \bibinfo
  {pages} {185} (\bibinfo {year} {2014})},\ \Eprint
  {http://arxiv.org/abs/1407.7056} {arXiv:1407.7056} \BibitemShut {NoStop}%
\bibitem [{\citenamefont {{Robles}}\ \emph {et~al.}(2015)\citenamefont
  {{Robles}}, \citenamefont {{Lora}}, \citenamefont {{Matos}},\ and\
  \citenamefont {{S{\'a}nchez-Salcedo}}}]{Matos2015}%
  \BibitemOpen
  \bibfield  {author} {\bibinfo {author} {\bibfnamefont {V.~H.}\ \bibnamefont
  {{Robles}}}, \bibinfo {author} {\bibfnamefont {V.}~\bibnamefont {{Lora}}},
  \bibinfo {author} {\bibfnamefont {T.}~\bibnamefont {{Matos}}}, \ and\
  \bibinfo {author} {\bibfnamefont {F.~J.}\ \bibnamefont
  {{S{\'a}nchez-Salcedo}}},\ }\href {\doibase 10.1088/0004-637X/810/2/99}
  {\bibfield  {journal} {\bibinfo  {journal} {apj}\ }\textbf {\bibinfo {volume}
  {810}},\ \bibinfo {eid} {99} (\bibinfo {year} {2015})},\ \Eprint
  {http://arxiv.org/abs/1404.3424} {arXiv:1404.3424} \BibitemShut {NoStop}%
\bibitem [{\citenamefont {Frieman}\ \emph {et~al.}(1995)\citenamefont
  {Frieman}, \citenamefont {Hill}, \citenamefont {Stebbins},\ and\
  \citenamefont {Waga}}]{Frieman1995}%
  \BibitemOpen
  \bibfield  {author} {\bibinfo {author} {\bibfnamefont {J.~A.}\ \bibnamefont
  {Frieman}}, \bibinfo {author} {\bibfnamefont {C.~T.}\ \bibnamefont {Hill}},
  \bibinfo {author} {\bibfnamefont {A.}~\bibnamefont {Stebbins}}, \ and\
  \bibinfo {author} {\bibfnamefont {I.}~\bibnamefont {Waga}},\ }\href {\doibase
  10.1103/PhysRevLett.75.2077} {\bibfield  {journal} {\bibinfo  {journal}
  {Phys. Rev. Lett.}\ }\textbf {\bibinfo {volume} {75}},\ \bibinfo {pages}
  {2077} (\bibinfo {year} {1995})}\BibitemShut {NoStop}%
\bibitem [{\citenamefont {Matos}\ \emph {et~al.}(2000)\citenamefont {Matos},
  \citenamefont {Guzmán},\ and\ \citenamefont
  {Ure{\~{n}}a-L{\'{o}}pez}}]{Matos2000}%
  \BibitemOpen
  \bibfield  {author} {\bibinfo {author} {\bibfnamefont {T.}~\bibnamefont
  {Matos}}, \bibinfo {author} {\bibfnamefont {F.~S.}\ \bibnamefont {Guzmán}},
  \ and\ \bibinfo {author} {\bibfnamefont {L.~A.}\ \bibnamefont
  {Ure{\~{n}}a-L{\'{o}}pez}},\ }\href
  {http://stacks.iop.org/0264-9381/17/i=7/a=309} {\bibfield  {journal}
  {\bibinfo  {journal} {Classical and Quantum Gravity}\ }\textbf {\bibinfo
  {volume} {17}},\ \bibinfo {pages} {1707} (\bibinfo {year}
  {2000})}\BibitemShut {NoStop}%
\bibitem [{\citenamefont {Su{\'a}rez}\ \emph {et~al.}(2014)\citenamefont
  {Su{\'a}rez}, \citenamefont {Robles},\ and\ \citenamefont
  {Matos}}]{Suarez2014}%
  \BibitemOpen
  \bibfield  {author} {\bibinfo {author} {\bibfnamefont {A.}~\bibnamefont
  {Su{\'a}rez}}, \bibinfo {author} {\bibfnamefont {H.~V.}\ \bibnamefont
  {Robles}}, \ and\ \bibinfo {author} {\bibfnamefont {T.}~\bibnamefont
  {Matos}},\ }\enquote {\bibinfo {title} {A review on the scalar
  field/bose-einstein condensate dark matter model},}\ in\ \href {\doibase
  10.1007/978-3-319-02063-1_9} {\emph {\bibinfo {booktitle} {Accelerated Cosmic
  Expansion: Proceedings of the Fourth International Meeting on Gravitation and
  Cosmology}}},\ \bibinfo {editor} {edited by\ \bibinfo {editor} {\bibfnamefont
  {C.}~\bibnamefont {Moreno~Gonz{\'a}lez}}, \bibinfo {editor} {\bibfnamefont
  {E.~J.}\ \bibnamefont {Madriz~Aguilar}}, \ and\ \bibinfo {editor}
  {\bibfnamefont {M.~L.}\ \bibnamefont {Reyes~Barrera}}}\ (\bibinfo
  {publisher} {Springer International Publishing},\ \bibinfo {address} {Cham},\
  \bibinfo {year} {2014})\ pp.\ \bibinfo {pages} {107--142}\BibitemShut
  {NoStop}%
\bibitem [{\citenamefont {{Lee}}(2016)}]{Lee2016}%
  \BibitemOpen
  \bibfield  {author} {\bibinfo {author} {\bibfnamefont {J.-W.}\ \bibnamefont
  {{Lee}}},\ }\href {\doibase 10.1016/j.physletb.2016.03.016} {\bibfield
  {journal} {\bibinfo  {journal} {Physics Letters B}\ }\textbf {\bibinfo
  {volume} {756}},\ \bibinfo {pages} {166} (\bibinfo {year} {2016})},\ \Eprint
  {http://arxiv.org/abs/1511.06611} {arXiv:1511.06611} \BibitemShut {NoStop}%
\bibitem [{\citenamefont {{Schive}}\ \emph
  {et~al.}(2014{\natexlab{a}})\citenamefont {{Schive}}, \citenamefont
  {{Chiueh}},\ and\ \citenamefont {{Broadhurst}}}]{Schive2014a}%
  \BibitemOpen
  \bibfield  {author} {\bibinfo {author} {\bibfnamefont {H.-Y.}\ \bibnamefont
  {{Schive}}}, \bibinfo {author} {\bibfnamefont {T.}~\bibnamefont {{Chiueh}}},
  \ and\ \bibinfo {author} {\bibfnamefont {T.}~\bibnamefont {{Broadhurst}}},\
  }\href {\doibase 10.1038/nphys2996} {\bibfield  {journal} {\bibinfo
  {journal} {Nature Physics}\ }\textbf {\bibinfo {volume} {10}},\ \bibinfo
  {pages} {496} (\bibinfo {year} {2014}{\natexlab{a}})},\ \Eprint
  {http://arxiv.org/abs/1406.6586} {arXiv:1406.6586} \BibitemShut {NoStop}%
\bibitem [{\citenamefont {Ruffini}\ and\ \citenamefont
  {Bonazzola}(1969)}]{Ruffini}%
  \BibitemOpen
  \bibfield  {author} {\bibinfo {author} {\bibfnamefont {R.}~\bibnamefont
  {Ruffini}}\ and\ \bibinfo {author} {\bibfnamefont {S.}~\bibnamefont
  {Bonazzola}},\ }\href {\doibase 10.1103/PhysRev.187.1767} {\bibfield
  {journal} {\bibinfo  {journal} {Phys. Rev.}\ }\textbf {\bibinfo {volume}
  {187}},\ \bibinfo {pages} {1767} (\bibinfo {year} {1969})}\BibitemShut
  {NoStop}%
\bibitem [{\citenamefont {Guzm{\'{a}}n}\ and\ \citenamefont
  {Ure{\~{n}}a-L{\'{o}}pez}(2004)}]{Guzman2004}%
  \BibitemOpen
  \bibfield  {author} {\bibinfo {author} {\bibfnamefont {F.~S.}\ \bibnamefont
  {Guzm{\'{a}}n}}\ and\ \bibinfo {author} {\bibfnamefont {L.~A.}\ \bibnamefont
  {Ure{\~{n}}a-L{\'{o}}pez}},\ }\href {\doibase 10.1103/PhysRevD.69.124033}
  {\bibfield  {journal} {\bibinfo  {journal} {Physical Review D}\ }\textbf
  {\bibinfo {volume} {69}},\ \bibinfo {pages} {124033} (\bibinfo {year}
  {2004})},\ \Eprint {http://arxiv.org/abs/0404014} {arXiv:0404014 [gr-qc]}
  \BibitemShut {NoStop}%
\bibitem [{\citenamefont {Liebling}\ and\ \citenamefont
  {Palenzuela}(2012)}]{Liebling2012}%
  \BibitemOpen
  \bibfield  {author} {\bibinfo {author} {\bibfnamefont {S.~L.}\ \bibnamefont
  {Liebling}}\ and\ \bibinfo {author} {\bibfnamefont {C.}~\bibnamefont
  {Palenzuela}},\ }\href {\doibase 10.12942/lrr-2012-6} {\bibfield  {journal}
  {\bibinfo  {journal} {Living Reviews in Relativity}\ }\textbf {\bibinfo
  {volume} {15}},\ \bibinfo {pages} {1} (\bibinfo {year} {2012})},\ \Eprint
  {http://arxiv.org/abs/1202.5809} {arXiv:1202.5809} \BibitemShut {NoStop}%
\bibitem [{\citenamefont {{Baldeschi}}\ \emph {et~al.}(1983)\citenamefont
  {{Baldeschi}}, \citenamefont {{Gelmini}},\ and\ \citenamefont
  {{Ruffini}}}]{Baldeschi1983}%
  \BibitemOpen
  \bibfield  {author} {\bibinfo {author} {\bibfnamefont {M.~R.}\ \bibnamefont
  {{Baldeschi}}}, \bibinfo {author} {\bibfnamefont {G.~B.}\ \bibnamefont
  {{Gelmini}}}, \ and\ \bibinfo {author} {\bibfnamefont {R.}~\bibnamefont
  {{Ruffini}}},\ }\href {\doibase 10.1016/0370-2693(83)90688-3} {\bibfield
  {journal} {\bibinfo  {journal} {Physics Letters B}\ }\textbf {\bibinfo
  {volume} {122}},\ \bibinfo {pages} {221} (\bibinfo {year}
  {1983})}\BibitemShut {NoStop}%
\bibitem [{\citenamefont {Sin}(1994)}]{Sin1994}%
  \BibitemOpen
  \bibfield  {author} {\bibinfo {author} {\bibfnamefont {S.-J.}\ \bibnamefont
  {Sin}},\ }\href {\doibase 10.1103/PhysRevD.50.3650} {\bibfield  {journal}
  {\bibinfo  {journal} {Phys. Rev. D}\ }\textbf {\bibinfo {volume} {50}},\
  \bibinfo {pages} {3650} (\bibinfo {year} {1994})}\BibitemShut {NoStop}%
\bibitem [{\citenamefont {Rindler-Daller}\ and\ \citenamefont
  {Shapiro}(2012{\natexlab{a}})}]{RindlerDaller2012}%
  \BibitemOpen
  \bibfield  {author} {\bibinfo {author} {\bibfnamefont {T.}~\bibnamefont
  {Rindler-Daller}}\ and\ \bibinfo {author} {\bibfnamefont {P.~R.}\
  \bibnamefont {Shapiro}},\ }in\ \href
  {https://inspirehep.net/record/1184893/files/arXiv:1209.1835.pdf} {\emph
  {\bibinfo {booktitle} {{6th International Meeting on Gravitation and
  Cosmology Guadalajara, Jalisco, Mexico, May 21-25, 2012}}}}\ (\bibinfo {year}
  {2012})\ \Eprint {http://arxiv.org/abs/1209.1835} {arXiv:1209.1835
  [astro-ph.CO]} \BibitemShut {NoStop}%
%%CITATION = ARXIV:1209.1835;%%
\bibitem [{\citenamefont {Paredes}\ and\ \citenamefont
  {Michinel}(2016)}]{Paredes2016}%
  \BibitemOpen
  \bibfield  {author} {\bibinfo {author} {\bibfnamefont {A.}~\bibnamefont
  {Paredes}}\ and\ \bibinfo {author} {\bibfnamefont {H.}~\bibnamefont
  {Michinel}},\ }\href {\doibase 10.1016/j.dark.2016.02.003} {\bibfield
  {journal} {\bibinfo  {journal} {Phys. Dark Univ.}\ }\textbf {\bibinfo
  {volume} {12}},\ \bibinfo {pages} {50} (\bibinfo {year} {2016})},\ \Eprint
  {http://arxiv.org/abs/1512.05121} {arXiv:1512.05121 [astro-ph.CO]}
  \BibitemShut {NoStop}%
%%CITATION = ARXIV:1512.05121;%%
\bibitem [{\citenamefont {{Schive}}\ \emph
  {et~al.}(2014{\natexlab{b}})\citenamefont {{Schive}}, \citenamefont {{Liao}},
  \citenamefont {{Woo}}, \citenamefont {{Wong}}, \citenamefont {{Chiueh}},
  \citenamefont {{Broadhurst}},\ and\ \citenamefont {{Hwang}}}]{Schive2014b}%
  \BibitemOpen
  \bibfield  {author} {\bibinfo {author} {\bibfnamefont {H.-Y.}\ \bibnamefont
  {{Schive}}}, \bibinfo {author} {\bibfnamefont {M.-H.}\ \bibnamefont
  {{Liao}}}, \bibinfo {author} {\bibfnamefont {T.-P.}\ \bibnamefont {{Woo}}},
  \bibinfo {author} {\bibfnamefont {S.-K.}\ \bibnamefont {{Wong}}}, \bibinfo
  {author} {\bibfnamefont {T.}~\bibnamefont {{Chiueh}}}, \bibinfo {author}
  {\bibfnamefont {T.}~\bibnamefont {{Broadhurst}}}, \ and\ \bibinfo {author}
  {\bibfnamefont {W.-Y.~P.}\ \bibnamefont {{Hwang}}},\ }\href {\doibase
  10.1103/PhysRevLett.113.261302} {\bibfield  {journal} {\bibinfo  {journal}
  {Phys. Rev. Lett.}\ }\textbf {\bibinfo {volume} {113}},\ \bibinfo {eid}
  {261302} (\bibinfo {year} {2014}{\natexlab{b}})},\ \Eprint
  {http://arxiv.org/abs/1407.7762} {arXiv:1407.7762} \BibitemShut {NoStop}%
\bibitem [{\citenamefont {{Calabrese}}\ and\ \citenamefont
  {{Spergel}}(2016)}]{Calabrese2016}%
  \BibitemOpen
  \bibfield  {author} {\bibinfo {author} {\bibfnamefont {E.}~\bibnamefont
  {{Calabrese}}}\ and\ \bibinfo {author} {\bibfnamefont {D.~N.}\ \bibnamefont
  {{Spergel}}},\ }\href@noop {} {\bibfield  {journal} {\bibinfo  {journal}
  {ArXiv e-prints}\ } (\bibinfo {year} {2016})},\ \Eprint
  {http://arxiv.org/abs/1603.07321} {arXiv:1603.07321} \BibitemShut {NoStop}%
\bibitem [{\citenamefont {{Carrasco}}\ \emph {et~al.}(2010)\citenamefont
  {{Carrasco}}, \citenamefont {{Gomez}}, \citenamefont {{Verdugo}},
  \citenamefont {{Lee}}, \citenamefont {{Diaz}}, \citenamefont {{Bergmann}},
  \citenamefont {{Turner}}, \citenamefont {{Miller}},\ and\ \citenamefont
  {{West}}}]{Carrasco2010}%
  \BibitemOpen
  \bibfield  {author} {\bibinfo {author} {\bibfnamefont {E.~R.}\ \bibnamefont
  {{Carrasco}}}, \bibinfo {author} {\bibfnamefont {P.~L.}\ \bibnamefont
  {{Gomez}}}, \bibinfo {author} {\bibfnamefont {T.}~\bibnamefont {{Verdugo}}},
  \bibinfo {author} {\bibfnamefont {H.}~\bibnamefont {{Lee}}}, \bibinfo
  {author} {\bibfnamefont {R.}~\bibnamefont {{Diaz}}}, \bibinfo {author}
  {\bibfnamefont {M.}~\bibnamefont {{Bergmann}}}, \bibinfo {author}
  {\bibfnamefont {J.~E.~H.}\ \bibnamefont {{Turner}}}, \bibinfo {author}
  {\bibfnamefont {B.~W.}\ \bibnamefont {{Miller}}}, \ and\ \bibinfo {author}
  {\bibfnamefont {M.~J.}\ \bibnamefont {{West}}},\ }\href {\doibase
  10.1088/2041-8205/715/2/L160} {\bibfield  {journal} {\bibinfo  {journal}
  {Astrophys. J. Lett.}\ }\textbf {\bibinfo {volume} {715}},\ \bibinfo {pages}
  {L160} (\bibinfo {year} {2010})},\ \Eprint {http://arxiv.org/abs/1004.5410}
  {arXiv:1004.5410 [astro-ph.CO]} \BibitemShut {NoStop}%
\bibitem [{\citenamefont {{Massey}}\ \emph {et~al.}(2015)\citenamefont
  {{Massey}}, \citenamefont {{Williams}}, \citenamefont {{Smit}}, \citenamefont
  {{Swinbank}}, \citenamefont {{Kitching}}, \citenamefont {{Harvey}},
  \citenamefont {{Jauzac}}, \citenamefont {{Israel}}, \citenamefont {{Clowe}},
  \citenamefont {{Edge}}, \citenamefont {{Hilton}}, \citenamefont {{Jullo}},
  \citenamefont {{Leonard}}, \citenamefont {{Liesenborgs}}, \citenamefont
  {{Merten}}, \citenamefont {{Mohammed}}, \citenamefont {{Nagai}},
  \citenamefont {{Richard}}, \citenamefont {{Robertson}}, \citenamefont
  {{Saha}}, \citenamefont {{Santana}}, \citenamefont {{Stott}},\ and\
  \citenamefont {{Tittley}}}]{Massey2015}%
  \BibitemOpen
  \bibfield  {author} {\bibinfo {author} {\bibfnamefont {R.}~\bibnamefont
  {{Massey}}}, \bibinfo {author} {\bibfnamefont {L.}~\bibnamefont
  {{Williams}}}, \bibinfo {author} {\bibfnamefont {R.}~\bibnamefont {{Smit}}},
  \bibinfo {author} {\bibfnamefont {M.}~\bibnamefont {{Swinbank}}}, \bibinfo
  {author} {\bibfnamefont {T.~D.}\ \bibnamefont {{Kitching}}}, \bibinfo
  {author} {\bibfnamefont {D.}~\bibnamefont {{Harvey}}}, \bibinfo {author}
  {\bibfnamefont {M.}~\bibnamefont {{Jauzac}}}, \bibinfo {author}
  {\bibfnamefont {H.}~\bibnamefont {{Israel}}}, \bibinfo {author}
  {\bibfnamefont {D.}~\bibnamefont {{Clowe}}}, \bibinfo {author} {\bibfnamefont
  {A.}~\bibnamefont {{Edge}}}, \bibinfo {author} {\bibfnamefont
  {M.}~\bibnamefont {{Hilton}}}, \bibinfo {author} {\bibfnamefont
  {E.}~\bibnamefont {{Jullo}}}, \bibinfo {author} {\bibfnamefont
  {A.}~\bibnamefont {{Leonard}}}, \bibinfo {author} {\bibfnamefont
  {J.}~\bibnamefont {{Liesenborgs}}}, \bibinfo {author} {\bibfnamefont
  {J.}~\bibnamefont {{Merten}}}, \bibinfo {author} {\bibfnamefont
  {I.}~\bibnamefont {{Mohammed}}}, \bibinfo {author} {\bibfnamefont
  {D.}~\bibnamefont {{Nagai}}}, \bibinfo {author} {\bibfnamefont
  {J.}~\bibnamefont {{Richard}}}, \bibinfo {author} {\bibfnamefont
  {A.}~\bibnamefont {{Robertson}}}, \bibinfo {author} {\bibfnamefont
  {P.}~\bibnamefont {{Saha}}}, \bibinfo {author} {\bibfnamefont
  {R.}~\bibnamefont {{Santana}}}, \bibinfo {author} {\bibfnamefont
  {J.}~\bibnamefont {{Stott}}}, \ and\ \bibinfo {author} {\bibfnamefont
  {E.}~\bibnamefont {{Tittley}}},\ }\href {\doibase 10.1093/mnras/stv467}
  {\bibfield  {journal} {\bibinfo  {journal} {Mon. Not. Roy. Astron. Soc.}\
  }\textbf {\bibinfo {volume} {449}},\ \bibinfo {pages} {3393} (\bibinfo {year}
  {2015})},\ \Eprint {http://arxiv.org/abs/1504.03388} {arXiv:1504.03388}
  \BibitemShut {NoStop}%
\bibitem [{\citenamefont {Seidel}\ and\ \citenamefont
  {Suen}(1994)}]{Seidel1994}%
  \BibitemOpen
  \bibfield  {author} {\bibinfo {author} {\bibfnamefont {E.}~\bibnamefont
  {Seidel}}\ and\ \bibinfo {author} {\bibfnamefont {W.-M.}\ \bibnamefont
  {Suen}},\ }\href {http://link.aps.org/doi/10.1103/PhysRevLett.72.2516}
  {\bibfield  {journal} {\bibinfo  {journal} {Phys. Rev. Lett.}\ }\textbf
  {\bibinfo {volume} {72}},\ \bibinfo {pages} {2516} (\bibinfo {year}
  {1994})}\BibitemShut {NoStop}%
\bibitem [{\citenamefont {Guzmán}\ and\ \citenamefont
  {Ure{\~{n}}a-L{\'{o}}pez}(2006)}]{Guzman2006}%
  \BibitemOpen
  \bibfield  {author} {\bibinfo {author} {\bibfnamefont {F.~S.}\ \bibnamefont
  {Guzmán}}\ and\ \bibinfo {author} {\bibfnamefont {L.~A.}\ \bibnamefont
  {Ure{\~{n}}a-L{\'{o}}pez}},\ }\href
  {http://stacks.iop.org/0004-637X/645/i=2/a=814} {\bibfield  {journal}
  {\bibinfo  {journal} {Astrophys. J.}\ }\textbf {\bibinfo {volume} {645}},\
  \bibinfo {pages} {814} (\bibinfo {year} {2006})}\BibitemShut {NoStop}%
\bibitem [{\citenamefont {Bernal}\ and\ \citenamefont
  {Guzm\'an}(2006)}]{Bernal2006}%
  \BibitemOpen
  \bibfield  {author} {\bibinfo {author} {\bibfnamefont {A.}~\bibnamefont
  {Bernal}}\ and\ \bibinfo {author} {\bibfnamefont {F.~S.}\ \bibnamefont
  {Guzm\'an}},\ }\href {\doibase 10.1103/PhysRevD.74.103002} {\bibfield
  {journal} {\bibinfo  {journal} {Phys. Rev. D}\ }\textbf {\bibinfo {volume}
  {74}},\ \bibinfo {pages} {103002} (\bibinfo {year} {2006})}\BibitemShut
  {NoStop}%
\bibitem [{\citenamefont {Gonzalez}\ and\ \citenamefont
  {Guzman}(2011)}]{Gonzalez2011}%
  \BibitemOpen
  \bibfield  {author} {\bibinfo {author} {\bibfnamefont {J.~A.}\ \bibnamefont
  {Gonzalez}}\ and\ \bibinfo {author} {\bibfnamefont {F.~S.}\ \bibnamefont
  {Guzman}},\ }\href {\doibase 10.1103/PhysRevD.83.103513} {\bibfield
  {journal} {\bibinfo  {journal} {Phys. Rev.}\ }\textbf {\bibinfo {volume}
  {D83}},\ \bibinfo {pages} {103513} (\bibinfo {year} {2011})},\ \Eprint
  {http://arxiv.org/abs/1105.2066} {arXiv:1105.2066 [astro-ph.CO]} \BibitemShut
  {NoStop}%
%%CITATION = ARXIV:1105.2066;%%
\bibitem [{Note1()}]{Note1}%
  \BibitemOpen
  \bibinfo {note} {To emphasize that this work is focused on galactic rather
  than stellar scales, we will refer to these solutions as (solitonic) cores
  instead of boson stars. All of our results, however, are independent of this
  interpretation.}\BibitemShut {Stop}%
\bibitem [{\citenamefont {Ji}\ and\ \citenamefont {Sin}(1994)}]{Ji1994}%
  \BibitemOpen
  \bibfield  {author} {\bibinfo {author} {\bibfnamefont {S.~U.}\ \bibnamefont
  {Ji}}\ and\ \bibinfo {author} {\bibfnamefont {S.~J.}\ \bibnamefont {Sin}},\
  }\href {\doibase 10.1103/PhysRevD.50.3655} {\bibfield  {journal} {\bibinfo
  {journal} {Phys. Rev.}\ }\textbf {\bibinfo {volume} {D50}},\ \bibinfo {pages}
  {3655} (\bibinfo {year} {1994})},\ \Eprint
  {http://arxiv.org/abs/hep-ph/9409267} {arXiv:hep-ph/9409267 [hep-ph]}
  \BibitemShut {NoStop}%
%%CITATION = HEP-PH/9409267;%%
\bibitem [{\citenamefont {{Almgren}}\ \emph {et~al.}(2013)\citenamefont
  {{Almgren}}, \citenamefont {{Bell}}, \citenamefont {{Lijewski}},
  \citenamefont {{Luki{\'c}}},\ and\ \citenamefont {{Van
  Andel}}}]{Almgren2013}%
  \BibitemOpen
  \bibfield  {author} {\bibinfo {author} {\bibfnamefont {A.~S.}\ \bibnamefont
  {{Almgren}}}, \bibinfo {author} {\bibfnamefont {J.~B.}\ \bibnamefont
  {{Bell}}}, \bibinfo {author} {\bibfnamefont {M.~J.}\ \bibnamefont
  {{Lijewski}}}, \bibinfo {author} {\bibfnamefont {Z.}~\bibnamefont
  {{Luki{\'c}}}}, \ and\ \bibinfo {author} {\bibfnamefont {E.}~\bibnamefont
  {{Van Andel}}},\ }\href {\doibase 10.1088/0004-637X/765/1/39} {\bibfield
  {journal} {\bibinfo  {journal} {Astrophys. J.}\ }\textbf {\bibinfo {volume}
  {765}},\ \bibinfo {eid} {39} (\bibinfo {year} {2013})},\ \Eprint
  {http://arxiv.org/abs/1301.4498} {arXiv:1301.4498 [astro-ph.IM]} \BibitemShut
  {NoStop}%
\bibitem [{\citenamefont {Madelung}(1927)}]{Madelung1927}%
  \BibitemOpen
  \bibfield  {author} {\bibinfo {author} {\bibfnamefont {E.}~\bibnamefont
  {Madelung}},\ }\href {\doibase 10.1007/BF01400372} {\bibfield  {journal}
  {\bibinfo  {journal} {Zeitschrift für Physik}\ }\textbf {\bibinfo {volume}
  {40}},\ \bibinfo {pages} {322} (\bibinfo {year} {1927})}\BibitemShut
  {NoStop}%
\bibitem [{\citenamefont {{Adeel Ajaib}}(2013)}]{Ajaib2013}%
  \BibitemOpen
  \bibfield  {author} {\bibinfo {author} {\bibfnamefont {M.}~\bibnamefont
  {{Adeel Ajaib}}},\ }\href@noop {} {\bibfield  {journal} {\bibinfo  {journal}
  {ArXiv e-prints}\ } (\bibinfo {year} {2013})},\ \Eprint
  {http://arxiv.org/abs/1302.5601} {arXiv:1302.5601 [physics.comp-ph]}
  \BibitemShut {NoStop}%
\bibitem [{\citenamefont {{Turk}}\ \emph {et~al.}(2011)\citenamefont {{Turk}},
  \citenamefont {{Smith}}, \citenamefont {{Oishi}}, \citenamefont {{Skory}},
  \citenamefont {{Skillman}}, \citenamefont {{Abel}},\ and\ \citenamefont
  {{Norman}}}]{Turk2011}%
  \BibitemOpen
  \bibfield  {author} {\bibinfo {author} {\bibfnamefont {M.~J.}\ \bibnamefont
  {{Turk}}}, \bibinfo {author} {\bibfnamefont {B.~D.}\ \bibnamefont {{Smith}}},
  \bibinfo {author} {\bibfnamefont {J.~S.}\ \bibnamefont {{Oishi}}}, \bibinfo
  {author} {\bibfnamefont {S.}~\bibnamefont {{Skory}}}, \bibinfo {author}
  {\bibfnamefont {S.~W.}\ \bibnamefont {{Skillman}}}, \bibinfo {author}
  {\bibfnamefont {T.}~\bibnamefont {{Abel}}}, \ and\ \bibinfo {author}
  {\bibfnamefont {M.~L.}\ \bibnamefont {{Norman}}},\ }\href {\doibase
  10.1088/0067-0049/192/1/9} {\bibfield  {journal} {\bibinfo  {journal} {ApJS}\
  }\textbf {\bibinfo {volume} {192}},\ \bibinfo {pages} {9} (\bibinfo {year}
  {2011})},\ \Eprint {http://arxiv.org/abs/1011.3514} {arXiv:1011.3514
  [astro-ph.IM]} \BibitemShut {NoStop}%
\bibitem [{\citenamefont {Naraschewski}\ \emph {et~al.}(1996)\citenamefont
  {Naraschewski}, \citenamefont {Wallis}, \citenamefont {Schenzle},
  \citenamefont {Cirac},\ and\ \citenamefont {Zoller}}]{Naraschewski1996}%
  \BibitemOpen
  \bibfield  {author} {\bibinfo {author} {\bibfnamefont {M.}~\bibnamefont
  {Naraschewski}}, \bibinfo {author} {\bibfnamefont {H.}~\bibnamefont
  {Wallis}}, \bibinfo {author} {\bibfnamefont {A.}~\bibnamefont {Schenzle}},
  \bibinfo {author} {\bibfnamefont {J.~I.}\ \bibnamefont {Cirac}}, \ and\
  \bibinfo {author} {\bibfnamefont {P.}~\bibnamefont {Zoller}},\ }\href
  {\doibase 10.1103/PhysRevA.54.2185} {\bibfield  {journal} {\bibinfo
  {journal} {Phys. Rev. A}\ }\textbf {\bibinfo {volume} {54}},\ \bibinfo
  {pages} {2185} (\bibinfo {year} {1996})}\BibitemShut {NoStop}%
\bibitem [{\citenamefont {R\"ohrl}\ \emph {et~al.}(1997)\citenamefont
  {R\"ohrl}, \citenamefont {Naraschewski}, \citenamefont {Schenzle},\ and\
  \citenamefont {Wallis}}]{Roehrl1997}%
  \BibitemOpen
  \bibfield  {author} {\bibinfo {author} {\bibfnamefont {A.}~\bibnamefont
  {R\"ohrl}}, \bibinfo {author} {\bibfnamefont {M.}~\bibnamefont
  {Naraschewski}}, \bibinfo {author} {\bibfnamefont {A.}~\bibnamefont
  {Schenzle}}, \ and\ \bibinfo {author} {\bibfnamefont {H.}~\bibnamefont
  {Wallis}},\ }\href {\doibase 10.1103/PhysRevLett.78.4143} {\bibfield
  {journal} {\bibinfo  {journal} {Phys. Rev. Lett.}\ }\textbf {\bibinfo
  {volume} {78}},\ \bibinfo {pages} {4143} (\bibinfo {year}
  {1997})}\BibitemShut {NoStop}%
\bibitem [{\citenamefont {Bernal}\ and\ \citenamefont
  {Guzman}(2006)}]{Bernal2006b}%
  \BibitemOpen
  \bibfield  {author} {\bibinfo {author} {\bibfnamefont {A.}~\bibnamefont
  {Bernal}}\ and\ \bibinfo {author} {\bibfnamefont {F.~S.}\ \bibnamefont
  {Guzman}},\ }\href {\doibase 10.1103/PhysRevD.74.063504} {\bibfield
  {journal} {\bibinfo  {journal} {Phys. Rev.}\ }\textbf {\bibinfo {volume}
  {D74}},\ \bibinfo {pages} {063504} (\bibinfo {year} {2006})},\ \Eprint
  {http://arxiv.org/abs/astro-ph/0608523} {arXiv:astro-ph/0608523 [astro-ph]}
  \BibitemShut {NoStop}%
%%CITATION = ASTRO-PH/0608523;%%
\bibitem [{\citenamefont {Rindler-Daller}\ and\ \citenamefont
  {Shapiro}(2010)}]{RindlerDaller2009}%
  \BibitemOpen
  \bibfield  {author} {\bibinfo {author} {\bibfnamefont {T.}~\bibnamefont
  {Rindler-Daller}}\ and\ \bibinfo {author} {\bibfnamefont {P.~R.}\
  \bibnamefont {Shapiro}},\ }\bibfield  {booktitle} {\emph {\bibinfo
  {booktitle} {{New Horizons in Astronomy: Frank N. Bash Symposium 2009, ASP
  Conf. Series Vol. 432 (2010), p.244, San Francisco: Astronomical Society of
  the Pacific}}},\ }\href@noop {} {\bibfield  {journal} {\bibinfo  {journal}
  {ASP Conf. Ser.}\ }\textbf {\bibinfo {volume} {432}},\ \bibinfo {pages} {244}
  (\bibinfo {year} {2010})},\ \Eprint {http://arxiv.org/abs/0912.2897}
  {arXiv:0912.2897 [astro-ph.CO]} \BibitemShut {NoStop}%
%%CITATION = ARXIV:0912.2897;%%
\bibitem [{\citenamefont {Rindler-Daller}\ and\ \citenamefont
  {Shapiro}(2012{\natexlab{b}})}]{RindlerDaller2011}%
  \BibitemOpen
  \bibfield  {author} {\bibinfo {author} {\bibfnamefont {T.}~\bibnamefont
  {Rindler-Daller}}\ and\ \bibinfo {author} {\bibfnamefont {P.~R.}\
  \bibnamefont {Shapiro}},\ }\href {\doibase 10.1111/j.1365-2966.2012.20588.x}
  {\bibfield  {journal} {\bibinfo  {journal} {Mon. Not. Roy. Astron. Soc.}\
  }\textbf {\bibinfo {volume} {422}},\ \bibinfo {pages} {135} (\bibinfo {year}
  {2012}{\natexlab{b}})},\ \Eprint {http://arxiv.org/abs/1106.1256}
  {arXiv:1106.1256 [astro-ph.CO]} \BibitemShut {NoStop}%
%%CITATION = ARXIV:1106.1256;%%
\bibitem [{\citenamefont {Li}\ \emph {et~al.}(2014)\citenamefont {Li},
  \citenamefont {Rindler-Daller},\ and\ \citenamefont
  {Shapiro}}]{RindlerDaller2013a}%
  \BibitemOpen
  \bibfield  {author} {\bibinfo {author} {\bibfnamefont {B.}~\bibnamefont
  {Li}}, \bibinfo {author} {\bibfnamefont {T.}~\bibnamefont {Rindler-Daller}},
  \ and\ \bibinfo {author} {\bibfnamefont {P.~R.}\ \bibnamefont {Shapiro}},\
  }\href {\doibase 10.1103/PhysRevD.89.083536} {\bibfield  {journal} {\bibinfo
  {journal} {Phys. Rev.}\ }\textbf {\bibinfo {volume} {D89}},\ \bibinfo {pages}
  {083536} (\bibinfo {year} {2014})},\ \Eprint {http://arxiv.org/abs/1310.6061}
  {arXiv:1310.6061 [astro-ph.CO]} \BibitemShut {NoStop}%
%%CITATION = ARXIV:1310.6061;%%
\bibitem [{\citenamefont {Rindler-Daller}\ and\ \citenamefont
  {Shapiro}(2014)}]{RindlerDaller2013b}%
  \BibitemOpen
  \bibfield  {author} {\bibinfo {author} {\bibfnamefont {T.}~\bibnamefont
  {Rindler-Daller}}\ and\ \bibinfo {author} {\bibfnamefont {P.~R.}\
  \bibnamefont {Shapiro}},\ }\href {\doibase 10.1142/S021773231430002X}
  {\bibfield  {journal} {\bibinfo  {journal} {Mod. Phys. Lett.}\ }\textbf
  {\bibinfo {volume} {A29}},\ \bibinfo {pages} {1430002} (\bibinfo {year}
  {2014})},\ \Eprint {http://arxiv.org/abs/1312.1734} {arXiv:1312.1734
  [astro-ph.CO]} \BibitemShut {NoStop}%
%%CITATION = ARXIV:1312.1734;%%
\bibitem [{\citenamefont {{Guzm{\'a}n}}\ \emph {et~al.}(2016)\citenamefont
  {{Guzm{\'a}n}}, \citenamefont {{Gonz{\'a}lez}},\ and\ \citenamefont
  {{Cruz-P{\'e}rez}}}]{Guzman2016}%
  \BibitemOpen
  \bibfield  {author} {\bibinfo {author} {\bibfnamefont {F.~S.}\ \bibnamefont
  {{Guzm{\'a}n}}}, \bibinfo {author} {\bibfnamefont {J.~A.}\ \bibnamefont
  {{Gonz{\'a}lez}}}, \ and\ \bibinfo {author} {\bibfnamefont {J.~P.}\
  \bibnamefont {{Cruz-P{\'e}rez}}},\ }\href {\doibase
  10.1103/PhysRevD.93.103535} {\bibfield  {journal} {\bibinfo  {journal}
  {\prd}\ }\textbf {\bibinfo {volume} {93}},\ \bibinfo {eid} {103535} (\bibinfo
  {year} {2016})},\ \Eprint {http://arxiv.org/abs/1605.04856}
  {arXiv:1605.04856} \BibitemShut {NoStop}%
\bibitem [{\citenamefont {{Du}}\ \emph {et~al.}(2016)\citenamefont {{Du}},
  \citenamefont {{Behrens}},\ and\ \citenamefont {{Niemeyer}}}]{Du2016}%
  \BibitemOpen
  \bibfield  {author} {\bibinfo {author} {\bibfnamefont {X.}~\bibnamefont
  {{Du}}}, \bibinfo {author} {\bibfnamefont {C.}~\bibnamefont {{Behrens}}}, \
  and\ \bibinfo {author} {\bibfnamefont {J.}~\bibnamefont {{Niemeyer}}},\
  }\href@noop {} {\bibfield  {journal} {\bibinfo  {journal} {in preparation}\ }
  (\bibinfo {year} {2016})}\BibitemShut {NoStop}%
\bibitem [{\citenamefont {{Schive}}\ \emph {et~al.}(2016)\citenamefont
  {{Schive}}, \citenamefont {{Chiueh}}, \citenamefont {{Broadhurst}},\ and\
  \citenamefont {{Huang}}}]{Schive2016}%
  \BibitemOpen
  \bibfield  {author} {\bibinfo {author} {\bibfnamefont {H.-Y.}\ \bibnamefont
  {{Schive}}}, \bibinfo {author} {\bibfnamefont {T.}~\bibnamefont {{Chiueh}}},
  \bibinfo {author} {\bibfnamefont {T.}~\bibnamefont {{Broadhurst}}}, \ and\
  \bibinfo {author} {\bibfnamefont {K.-W.}\ \bibnamefont {{Huang}}},\ }\href
  {\doibase 10.3847/0004-637X/818/1/89} {\bibfield  {journal} {\bibinfo
  {journal} {Astrophys. J.}\ }\textbf {\bibinfo {volume} {818}},\ \bibinfo
  {eid} {89} (\bibinfo {year} {2016})},\ \Eprint
  {http://arxiv.org/abs/1508.04621} {arXiv:1508.04621} \BibitemShut {NoStop}%
\bibitem [{\citenamefont {{Sarkar}}\ \emph {et~al.}(2016)\citenamefont
  {{Sarkar}}, \citenamefont {{Mondal}}, \citenamefont {{Das}}, \citenamefont
  {{Sethi}}, \citenamefont {{Bharadwaj}},\ and\ \citenamefont
  {{Marsh}}}]{Sarkar2016}%
  \BibitemOpen
  \bibfield  {author} {\bibinfo {author} {\bibfnamefont {A.}~\bibnamefont
  {{Sarkar}}}, \bibinfo {author} {\bibfnamefont {R.}~\bibnamefont {{Mondal}}},
  \bibinfo {author} {\bibfnamefont {S.}~\bibnamefont {{Das}}}, \bibinfo
  {author} {\bibfnamefont {S.~K.}\ \bibnamefont {{Sethi}}}, \bibinfo {author}
  {\bibfnamefont {S.}~\bibnamefont {{Bharadwaj}}}, \ and\ \bibinfo {author}
  {\bibfnamefont {D.~J.~E.}\ \bibnamefont {{Marsh}}},\ }\href {\doibase
  10.1088/1475-7516/2016/04/012} {\bibfield  {journal} {\bibinfo  {journal}
  {JCAP}\ }\textbf {\bibinfo {volume} {4}},\ \bibinfo {eid} {012} (\bibinfo
  {year} {2016})},\ \Eprint {http://arxiv.org/abs/1512.03325}
  {arXiv:1512.03325} \BibitemShut {NoStop}%
\bibitem [{\citenamefont {{Marsh}}(2016)}]{Marsh2016}%
  \BibitemOpen
  \bibfield  {author} {\bibinfo {author} {\bibfnamefont {D.~J.~E.}\
  \bibnamefont {{Marsh}}},\ }\href@noop {} {\bibfield  {journal} {\bibinfo
  {journal} {ArXiv e-prints}\ } (\bibinfo {year} {2016})},\ \Eprint
  {http://arxiv.org/abs/1605.05973} {arXiv:1605.05973} \BibitemShut {NoStop}%
\end{thebibliography}


\end{document}